%% file: louis.tex
\documentstyle[twoside,fleqn,espcrc2]{article}

\newcommand{\AmS}{{\protect\the\textfont2
  A\kern-.1667em\lower.5ex\hbox{M}\kern-.125emS}}

\title{Correlation functions in lattice formulations of quantum 
gravity\thanks{Supported in part by Fonds zur F\"orderung der
        wissenschaftlichen Forschung under Contract P11141-PHY.}}

\author{W. Beirl, A. Hauke, P. Homolka, H. Markum and J. Riedler\address{
        Institut f\"ur Kernphysik, Technische Universit\"at Wien, 
        A-1040 Vienna, Austria}}
       
\begin{document}

\begin{abstract}
We compare different models of a quantum theory of four-dimensional
lattice gravity based on Regge's original proposal. From Monte Carlo
simulations we calculate two-point functions between geometrical quantities 
and estimate the masses of the corresponding interaction particles.
\end{abstract}

\maketitle

\section{INTRODUCTION}
There are two related schools attempting at a lattice theory of quantum
gravity, dynamical triangulation and Regge quantum gravity. We concentrate
on the latter approach and refer to \cite{cat,des} for excellent reviews of 
the former framework. In particular we investigate
different formulations, all of them based on the original work by Regge
\cite{reg}, namely conventional Regge gravity \cite{ber1,ham1}, a
group theoretical approach \cite{cas}, $Z_2$-link Regge-theory
\cite{juri}, and Regge gravity coupled to SU(2)-gauge theory \cite{ber2}.
Each model exposes phase transitions at some critical
gravitational couplings separating small and large curvature phases
and allows to look for a continuum limit.
\cite{pet,kri}. A candidate realistic quantum theory of gravity should
reproduce the expected long range interaction behavior observed in
nature. Thus we calculate two-point funtions to probe the existence of
massless quanta of the gravitational field.

\section{LATTICE QUANTUM GRAVITY}
Any smooth $d$-manifold can be approximated by appropriately glueing together
pieces of flat space, called $d$-simplices, ending up with a simplicial
lattice. We take the edge lengths as degrees of freedom and leave the 
triangulation of the lattice fixed. Adopting the Euclidean 
path integral we can write down the partition function
\begin{equation} \label{Z}
Z=\int D[q,U]e^{-I(q,U)}
\end{equation}
with an action $I$ that covers all of the models introduced in more detail 
below. The functional integration extends over the squared edge lengths 
$q$ and (if switched on) over the non-Abelian gauge fields $U$.

One of the problems with (\ref{Z}) is the ambiguity in performing the 
link-lengths integration. Commonly the measure is written as
\begin{equation} \label{Dq}
D[q]=\prod_l dq_lq_l^{\sigma-1}{\cal F}(q) ~,
\end{equation}
with ${\cal F}$ a function of the squared edge lengths $q$ being equal
to one if the Euclidean triangle inequalities are fulfilled and zero
otherwise. The question remains whether such a local measure is
sufficient and how the power $\sigma$ might be chosen. A recent
calculation for two-dimensional pure gravity pointed out that simple
measures like (\ref{Dq}) do not respect the infinite volume of the
diffeomorphism group. Only by appropriate gauge fixing and including 
the corresponding Faddeev-Popov term the correct continuum limit can 
be obtained \cite{men}. However, the generalization of this procedure 
to higher dimensions and its numerical implementation are
technically demanding.

Working in Euclidean space, i.e.~with positive definite metric,
the conformal mode renders the continuum Einstein-Hilbert action 
unbounded from below. This unpleasant feature persists in the
discretized Regge-Einstein action but need not necessarily lead to an
ill-defined path integral \cite{ber1}. Indeed, numerical simulations
reveal the existence of a well-defined phase with finite expectation
values within a certain range of the bare Planck mass, 
$(m_c^-)^2\le m_P^2\le (m_c^+)^2$ \cite{pet}. 

On a triangulated lattice an action is given by
\begin{eqnarray} \label{Iq}
I(q,U)\!\!\!&=\!\!\!&-2m_P^2\sum\limits_tR_t(A_t,\delta_t)+
\lambda\sum\limits_sV_s\nonumber\\
&&+\frac{\beta}{2}\sum\limits_tW_t\mbox{Re}[\mbox{Tr}(1-U_t)] ~.
\end{eqnarray}
The first sum runs over all triangle areas $A_t(q)$ and
corresponding deficit angles $\delta_t(q)$ yielding the curvature $R_t$.
A cosmological term consisting of the 
cosmological constant $\lambda$ times the sum over the volumes $V_s(q)$ 
of all four-simplices $s$ follows. Finally, an additional non-Abelian 
gauge action composed of the inverse gauge coupling $\beta$, the weight 
factors $W_t(q)$ and the ordered product $U_t$ of SU(2) matrices around 
the triangle $t$ is appended. The weights 
\begin{equation}
W_t=\mbox{const}\times\frac{V_t}{A_t^2} ~, 
\end{equation} 
with a four-volume $V_t$ assigned to every triangle, describe the coupling 
of gravity to the gauge fields. 

Now we are ready to define the four different models for subsequent
numerical treatment. Monte Carlo simulations have been 
performed on regularly triangulated hypercubic lattices with toroidal 
topology and $4^3\times 8$ vertices. The gravitational couplings $m_P^2$
were chosen close to $(m_c^-)^2$ where there is a certain chance for a
continuous phase transition \cite{pet}.

\subsection{Conventional Regge gravity}
Here we employ the Regge-Einstein action, putting $R_t=A_t\delta_t$
with the gravitational couplings $m_P^2=\{-0.0775,-0.0785,-0.0795\}$, and
include a cosmological term with $\lambda=1$ \cite{ber1,ham1}. No
additional gauge fields are present in the action, $\beta=0$, and the 
gravitational measure is chosen to be uniform, $\sigma=1$. 

\subsection{Group theoretical approach}
Constructing the dual of a simplicial lattice, Poincar{\'e} transformations 
can be assigned to its links to yield an action in which the {\em sin} 
of the deficit angle enters, \hbox{$R_t=A_t\sin\delta_t$} \cite{cas}. 
Again we use a cosmological constant, \hbox{$\lambda=1$}, vanishing inverse 
gauge coupling, \hbox{$\beta=0$}, the uniform measure, $\sigma=1$, and vary
\hbox{$m_P^2=\{-0.055,-0.0555,-0.056\}$}.

\subsection{$Z_2$-link Regge-gravity}
This model is defined by restricting the squared link lengths to take on
only two possible values
\begin{equation} \label{ql}
q_l\sim 1+\epsilon\sigma_l ~,\quad\sigma_l\in Z_2 ~, 
\quad\epsilon\le\epsilon_{max}\in I\!\! R_+ ~.
\end{equation}
Then the quantum-gravity path-integral can be rewritten as the partition
function of a spin system with somewhat complicated, yet local spin
interactions \cite{juri}. 

\subsection{Gauge fields coupled to Regge gravity}
In the system of SU(2) gauge fields coupled to quantum gravity we set
$R_t=A_t\delta_t$, \hbox{$\lambda=0$} and use a scale invariant 
measure, $\sigma=0$ \cite{ber2,kri}. The following pairs of 
gravitational and gauge coupling, respectively, are considered:
\hbox{$(m_P^2,\beta)=\{(-0.0025,1.6),(-0.005,1.0)\}$}.

\section{TWO-POINT FUNCTIONS}
In order to examine the physical relevance of the well-defined phase
mentioned above, we compute correlation functions of certain geometrical
quantities. The gravitational weak-field propagator 
can be cast into a spin-two and a spin-zero contribution
\cite{ham2}. The volume correlations are sensitive to the scalar part
\begin{equation} \label{GV}
G_V(d)=\langle\sum_{s\supset v_0}V_s\sum_{s'\supset v_d}V_{s'}\rangle_c
\end{equation}
and the curvature correlations are due to the presence of spin-2
particles
\begin{equation} \label{GR}
G_R(d)=\langle\sum_{t\supset v_0}R_t\sum_{t'\supset v_d}R_{t'}\rangle_c ~.
\end{equation}
The local operators in (\ref{GV}) and (\ref{GR}) should be measured at two
vertices $v_0$ and $v_d$ separated by the geodesic distance.
We take the distance $d$ to be equal to the index distance along the main 
axes of the skeleton. This seems a reasonable approximation in the 
well-defined phase with its small average curvature.
In general one expects for (\ref{GV}) and (\ref{GR}) at large distances
the functional form
\begin{equation} \label{G}
G\sim\frac{e^{-md}}{d^a}  ~.
\end{equation}
A power law with $a=2$ and a vanishing effective mass $m=0$ would hint
at Newtonian gravity with massless gravitons.

\begin{figure*}[t]
\hbox{(a) Conventional Regge gravity: $\quad\Diamond\dots m_P^2=-0.0775,
     \quad\Box\dots m_P^2=-0.0785, \quad\triangle\dots m_P^2=-0.0795$}
\vspace{3mm}
\hbox{\input{fig11.tex} \input{fig12.tex} \input{fig13.tex}}
\vspace{6mm}
\hbox{\hspace{1mm} (b) Group theoretical approach:
      $\quad\Diamond\dots m_P^2=-0.055, \quad\Box\dots m_P^2=-0.0555,
       \quad\triangle\dots m_P^2=-0.056$}
\vspace{3mm}
\hbox{\input{fig21.tex} \input{fig22.tex} \input{fig23.tex}}
\vspace{6mm}
\hbox{(c) Gauge fields coupled to gravity: $\quad\Diamond\dots
      (m_P^2=-0.0025,\beta=1.6), \quad\Box\dots
(m_P^2=-0.005,\beta=1.0)$}
\vspace{3mm}
\hbox{\input{fig31.tex} \input{fig32.tex} \input{fig33.tex}}
\caption{Volume (left plots) and curvature (middle with magnification in
the right plots) correlation functions for (a) conventional Regge
gravity,
(b) the group theoretical approach, and (c) the system of non-Abelian
gauge fields coupled to Regge gravity. Error bars not explicitely drawn
are in the size of the symbols. The curves correspond to fits with
the function (8).}
\label{Fig}
\end{figure*}
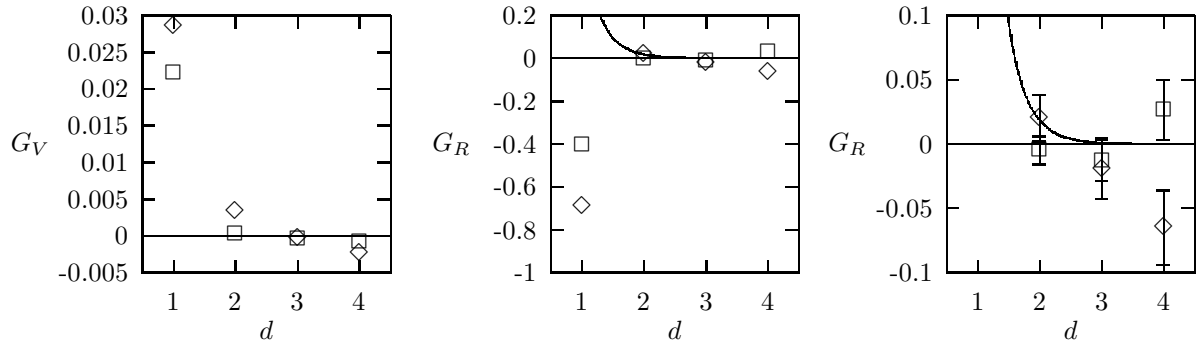
\clearpage
\noindent
Table 1 \\
\hbox{Effective masses $m$ of $G_R$ for several gravitational couplings
$m_P^2$} \\
\begin{tabular*}{\textwidth}{@{}l@{\extracolsep{\fill}}lllllll}
\hline
&\multicolumn{3}{c}{(a) Conventional Regge gravity}&\multicolumn{3}{c}
{(b) Group theoretical approach}&\multicolumn{1}{c}{(c) Gauge fields}\\
\cline{2-4} \cline{5-7} \cline{8-8}
$m_P^{2^{^{ }}}$~ & $-0.0775$ & $-0.0785$ & $-0.0795$ & $-0.055$
& $-0.0555$ & $-0.056$ & $-0.0025$ \\
$m$ & ~~$\,2.9$ & ~~$\,3.1$ & ~~$\,3.0$ & ~~$\,1.1$ & ~~$\,0.9$
& ~~$\,0.5$ & ~~$\,2.1$ \\
\hline
\end{tabular*}
\vspace*{5mm}

\section{RESULTS}
Fig.~1 displays our Monte Carlo data of the two-point functions (\ref{GV})
and (\ref{GR}). The volume correlations have already been studied around the
transition at positive coupling where fits to an exponential decay have
been obtained \cite{dal}. Such a fit procedure seems to be more
difficult at negative couplings.

The curvature correlations are more suitable for a fit with (\ref{G}). 
In order to test whether they obey a power law we fixed $a=2$ and fitted the 
effective masses $m$. We took only distances $d\ge 2$ into account.
$G_R(d=1)$ is presumably plagued by lattice artifacts due to
contact terms \cite{hom}. We are anyhow interested in the large distance 
behavior. 

Table~1 contains the obtained mass parameters $m$. For (a) conventional Regge
gravity $m$ stays rather constant towards the critical coupling
whereas in (b) the group theoretical approach the mass decreases for
$m_P^2\to (m_c^-)^2$. In the case of SU(2) fields on the fluctuating
lattice we get only one reasonable fit again indicating a nonzero mass.
For all fits the uncertainties in the mass parameters are large. The 
$Z_2$-link approximation of Regge gravity seems to behave
differently. For this model the curvature correlations are compatible with 
a power law. The results of a more extended study of the spin approach are 
reported in \cite{hom}.

\section{SUMMARY}
We computed two-point functions close to the critical bare Planck mass in 
the negative gravitational coupling regime. Altogether they exhibit a very 
similar behavior for the considered models. Except for $Z_2$-link
Regge-gravity we found no convincing evidence for long range 
correlations corresponding\, to\, massless\, spin-zero\, or\, spin-two
\newpage
\vspace*{29mm}
\noindent
excitations. One reason might be, that contrary to the models described here,
the $Z_2$-link theory is well-defined for all couplings $m_P^2$ and
computationally much less demanding, which allows 
to perform simulations exactly at the critical point.

\end{document}

%% file: fig11.tex
\setlength{\unitlength}{0.240900pt}
\ifx\plotpoint\undefined\newsavebox{\plotpoint}\fi
\sbox{\plotpoint}{\rule[-0.200pt]{0.400pt}{0.400pt}}%
\begin{picture}(600,540)(0,0)
\font\gnuplot=cmr10 at 10pt
\gnuplot
\sbox{\plotpoint}{\rule[-0.200pt]{0.400pt}{0.400pt}}%
\put(220.0,171.0){\rule[-0.200pt]{93.951pt}{0.400pt}}
\put(220.0,113.0){\rule[-0.200pt]{4.818pt}{0.400pt}}
\put(198,113){\makebox(0,0)[r]{-0.001}}
\put(590.0,113.0){\rule[-0.200pt]{4.818pt}{0.400pt}}
\put(220.0,171.0){\rule[-0.200pt]{4.818pt}{0.400pt}}
\put(198,171){\makebox(0,0)[r]{0}}
\put(590.0,171.0){\rule[-0.200pt]{4.818pt}{0.400pt}}
\put(220.0,228.0){\rule[-0.200pt]{4.818pt}{0.400pt}}
\put(198,228){\makebox(0,0)[r]{0.001}}
\put(590.0,228.0){\rule[-0.200pt]{4.818pt}{0.400pt}}
\put(220.0,286.0){\rule[-0.200pt]{4.818pt}{0.400pt}}
\put(198,286){\makebox(0,0)[r]{0.002}}
\put(590.0,286.0){\rule[-0.200pt]{4.818pt}{0.400pt}}
\put(220.0,344.0){\rule[-0.200pt]{4.818pt}{0.400pt}}
\put(198,344){\makebox(0,0)[r]{0.003}}
\put(590.0,344.0){\rule[-0.200pt]{4.818pt}{0.400pt}}
\put(220.0,402.0){\rule[-0.200pt]{4.818pt}{0.400pt}}
\put(198,402){\makebox(0,0)[r]{0.004}}
\put(590.0,402.0){\rule[-0.200pt]{4.818pt}{0.400pt}}
\put(220.0,459.0){\rule[-0.200pt]{4.818pt}{0.400pt}}
\put(198,459){\makebox(0,0)[r]{0.005}}
\put(590.0,459.0){\rule[-0.200pt]{4.818pt}{0.400pt}}
\put(220.0,517.0){\rule[-0.200pt]{4.818pt}{0.400pt}}
\put(198,517){\makebox(0,0)[r]{0.006}}
\put(590.0,517.0){\rule[-0.200pt]{4.818pt}{0.400pt}}
\put(269.0,113.0){\rule[-0.200pt]{0.400pt}{4.818pt}}
\put(269,68){\makebox(0,0){1}}
\put(269.0,497.0){\rule[-0.200pt]{0.400pt}{4.818pt}}
\put(366.0,113.0){\rule[-0.200pt]{0.400pt}{4.818pt}}
\put(366,68){\makebox(0,0){2}}
\put(366.0,497.0){\rule[-0.200pt]{0.400pt}{4.818pt}}
\put(464.0,113.0){\rule[-0.200pt]{0.400pt}{4.818pt}}
\put(464,68){\makebox(0,0){3}}
\put(464.0,497.0){\rule[-0.200pt]{0.400pt}{4.818pt}}
\put(561.0,113.0){\rule[-0.200pt]{0.400pt}{4.818pt}}
\put(561,68){\makebox(0,0){4}}
\put(561.0,497.0){\rule[-0.200pt]{0.400pt}{4.818pt}}
\put(220.0,113.0){\rule[-0.200pt]{93.951pt}{0.400pt}}
\put(610.0,113.0){\rule[-0.200pt]{0.400pt}{97.324pt}}
\put(220.0,517.0){\rule[-0.200pt]{93.951pt}{0.400pt}}
\put(45,315){\makebox(0,0){$G_V$}}
\put(415,23){\makebox(0,0){$d$}}
\put(220.0,113.0){\rule[-0.200pt]{0.400pt}{97.324pt}}
\put(269,453){\raisebox{-.8pt}{\makebox(0,0){$\Diamond$}}}
\put(366,161){\raisebox{-.8pt}{\makebox(0,0){$\Diamond$}}}
\put(464,168){\raisebox{-.8pt}{\makebox(0,0){$\Diamond$}}}
\put(561,173){\raisebox{-.8pt}{\makebox(0,0){$\Diamond$}}}
\put(269,451){\raisebox{-.8pt}{\makebox(0,0){$\Box$}}}
\put(366,167){\raisebox{-.8pt}{\makebox(0,0){$\Box$}}}
\put(464,167){\raisebox{-.8pt}{\makebox(0,0){$\Box$}}}
\put(561,165){\raisebox{-.8pt}{\makebox(0,0){$\Box$}}}
\put(269,452){\makebox(0,0){$\triangle$}}
\put(366,163){\makebox(0,0){$\triangle$}}
\put(464,156){\makebox(0,0){$\triangle$}}
\put(561,160){\makebox(0,0){$\triangle$}}
\end{picture}

%% file: fig12.tex
\setlength{\unitlength}{0.240900pt}
\ifx\plotpoint\undefined\newsavebox{\plotpoint}\fi
\sbox{\plotpoint}{\rule[-0.200pt]{0.400pt}{0.400pt}}%
\begin{picture}(580,540)(0,0)
\font\gnuplot=cmr10 at 10pt
\gnuplot
\sbox{\plotpoint}{\rule[-0.200pt]{0.400pt}{0.400pt}}%
\put(220.0,436.0){\rule[-0.200pt]{93.951pt}{0.400pt}}
\put(220.0,113.0){\rule[-0.200pt]{4.818pt}{0.400pt}}
\put(198,113){\makebox(0,0)[r]{-0.2}}
\put(590.0,113.0){\rule[-0.200pt]{4.818pt}{0.400pt}}
\put(220.0,194.0){\rule[-0.200pt]{4.818pt}{0.400pt}}
\put(198,194){\makebox(0,0)[r]{-0.15}}
\put(590.0,194.0){\rule[-0.200pt]{4.818pt}{0.400pt}}
\put(220.0,275.0){\rule[-0.200pt]{4.818pt}{0.400pt}}
\put(198,275){\makebox(0,0)[r]{-0.1}}
\put(590.0,275.0){\rule[-0.200pt]{4.818pt}{0.400pt}}
\put(220.0,355.0){\rule[-0.200pt]{4.818pt}{0.400pt}}
\put(198,355){\makebox(0,0)[r]{-0.05}}
\put(590.0,355.0){\rule[-0.200pt]{4.818pt}{0.400pt}}
\put(220.0,436.0){\rule[-0.200pt]{4.818pt}{0.400pt}}
\put(198,436){\makebox(0,0)[r]{0}}
\put(590.0,436.0){\rule[-0.200pt]{4.818pt}{0.400pt}}
\put(220.0,517.0){\rule[-0.200pt]{4.818pt}{0.400pt}}
\put(198,517){\makebox(0,0)[r]{0.05}}
\put(590.0,517.0){\rule[-0.200pt]{4.818pt}{0.400pt}}
\put(269.0,113.0){\rule[-0.200pt]{0.400pt}{4.818pt}}
\put(269,68){\makebox(0,0){1}}
\put(269.0,497.0){\rule[-0.200pt]{0.400pt}{4.818pt}}
\put(366.0,113.0){\rule[-0.200pt]{0.400pt}{4.818pt}}
\put(366,68){\makebox(0,0){2}}
\put(366.0,497.0){\rule[-0.200pt]{0.400pt}{4.818pt}}
\put(464.0,113.0){\rule[-0.200pt]{0.400pt}{4.818pt}}
\put(464,68){\makebox(0,0){3}}
\put(464.0,497.0){\rule[-0.200pt]{0.400pt}{4.818pt}}
\put(561.0,113.0){\rule[-0.200pt]{0.400pt}{4.818pt}}
\put(561,68){\makebox(0,0){4}}
\put(561.0,497.0){\rule[-0.200pt]{0.400pt}{4.818pt}}
\put(220.0,113.0){\rule[-0.200pt]{93.951pt}{0.400pt}}
\put(610.0,113.0){\rule[-0.200pt]{0.400pt}{97.324pt}}
\put(220.0,517.0){\rule[-0.200pt]{93.951pt}{0.400pt}}
\put(65,315){\makebox(0,0){$G_R$}}
\put(415,23){\makebox(0,0){$d$}}
\put(220.0,113.0){\rule[-0.200pt]{0.400pt}{97.324pt}}
\put(269,180){\raisebox{-.8pt}{\makebox(0,0){$\Diamond$}}}
\put(366,443){\raisebox{-.8pt}{\makebox(0,0){$\Diamond$}}}
\put(464,435){\raisebox{-.8pt}{\makebox(0,0){$\Diamond$}}}
\put(561,437){\raisebox{-.8pt}{\makebox(0,0){$\Diamond$}}}
\put(306,515.67){\rule{0.241pt}{0.400pt}}
\multiput(306.00,516.17)(0.500,-1.000){2}{\rule{0.120pt}{0.400pt}}
\multiput(307.60,510.19)(0.468,-1.797){5}{\rule{0.113pt}{1.400pt}}
\multiput(306.17,513.09)(4.000,-10.094){2}{\rule{0.400pt}{0.700pt}}
\multiput(311.60,498.02)(0.468,-1.505){5}{\rule{0.113pt}{1.200pt}}
\multiput(310.17,500.51)(4.000,-8.509){2}{\rule{0.400pt}{0.600pt}}
\multiput(315.61,486.60)(0.447,-1.802){3}{\rule{0.108pt}{1.300pt}}
\multiput(314.17,489.30)(3.000,-6.302){2}{\rule{0.400pt}{0.650pt}}
\multiput(318.60,479.68)(0.468,-0.920){5}{\rule{0.113pt}{0.800pt}}
\multiput(317.17,481.34)(4.000,-5.340){2}{\rule{0.400pt}{0.400pt}}
\multiput(322.60,473.09)(0.468,-0.774){5}{\rule{0.113pt}{0.700pt}}
\multiput(321.17,474.55)(4.000,-4.547){2}{\rule{0.400pt}{0.350pt}}
\multiput(326.60,467.09)(0.468,-0.774){5}{\rule{0.113pt}{0.700pt}}
\multiput(325.17,468.55)(4.000,-4.547){2}{\rule{0.400pt}{0.350pt}}
\multiput(330.00,462.94)(0.481,-0.468){5}{\rule{0.500pt}{0.113pt}}
\multiput(330.00,463.17)(2.962,-4.000){2}{\rule{0.250pt}{0.400pt}}
\multiput(334.00,458.94)(0.481,-0.468){5}{\rule{0.500pt}{0.113pt}}
\multiput(334.00,459.17)(2.962,-4.000){2}{\rule{0.250pt}{0.400pt}}
\multiput(338.00,454.95)(0.685,-0.447){3}{\rule{0.633pt}{0.108pt}}
\multiput(338.00,455.17)(2.685,-3.000){2}{\rule{0.317pt}{0.400pt}}
\put(342,451.17){\rule{0.900pt}{0.400pt}}
\multiput(342.00,452.17)(2.132,-2.000){2}{\rule{0.450pt}{0.400pt}}
\multiput(346.00,449.95)(0.685,-0.447){3}{\rule{0.633pt}{0.108pt}}
\multiput(346.00,450.17)(2.685,-3.000){2}{\rule{0.317pt}{0.400pt}}
\put(350,446.67){\rule{0.964pt}{0.400pt}}
\multiput(350.00,447.17)(2.000,-1.000){2}{\rule{0.482pt}{0.400pt}}
\put(354,445.17){\rule{0.900pt}{0.400pt}}
\multiput(354.00,446.17)(2.132,-2.000){2}{\rule{0.450pt}{0.400pt}}
\put(358,443.67){\rule{0.964pt}{0.400pt}}
\multiput(358.00,444.17)(2.000,-1.000){2}{\rule{0.482pt}{0.400pt}}
\put(362,442.67){\rule{0.964pt}{0.400pt}}
\multiput(362.00,443.17)(2.000,-1.000){2}{\rule{0.482pt}{0.400pt}}
\put(366,441.67){\rule{0.964pt}{0.400pt}}
\multiput(366.00,442.17)(2.000,-1.000){2}{\rule{0.482pt}{0.400pt}}
\put(370,440.67){\rule{0.964pt}{0.400pt}}
\multiput(370.00,441.17)(2.000,-1.000){2}{\rule{0.482pt}{0.400pt}}
\put(374,439.67){\rule{0.964pt}{0.400pt}}
\multiput(374.00,440.17)(2.000,-1.000){2}{\rule{0.482pt}{0.400pt}}
\put(382,438.67){\rule{0.723pt}{0.400pt}}
\multiput(382.00,439.17)(1.500,-1.000){2}{\rule{0.361pt}{0.400pt}}
\put(378.0,440.0){\rule[-0.200pt]{0.964pt}{0.400pt}}
\put(389,437.67){\rule{0.964pt}{0.400pt}}
\multiput(389.00,438.17)(2.000,-1.000){2}{\rule{0.482pt}{0.400pt}}
\put(385.0,439.0){\rule[-0.200pt]{0.964pt}{0.400pt}}
\put(405,436.67){\rule{0.964pt}{0.400pt}}
\multiput(405.00,437.17)(2.000,-1.000){2}{\rule{0.482pt}{0.400pt}}
\put(393.0,438.0){\rule[-0.200pt]{2.891pt}{0.400pt}}
\put(441,435.67){\rule{0.964pt}{0.400pt}}
\multiput(441.00,436.17)(2.000,-1.000){2}{\rule{0.482pt}{0.400pt}}
\put(409.0,437.0){\rule[-0.200pt]{7.709pt}{0.400pt}}
\put(445.0,436.0){\rule[-0.200pt]{39.748pt}{0.400pt}}
\put(269,193){\raisebox{-.8pt}{\makebox(0,0){$\Box$}}}
\put(366,443){\raisebox{-.8pt}{\makebox(0,0){$\Box$}}}
\put(464,434){\raisebox{-.8pt}{\makebox(0,0){$\Box$}}}
\put(561,436){\raisebox{-.8pt}{\makebox(0,0){$\Box$}}}
\put(309.67,515){\rule{0.400pt}{0.482pt}}
\multiput(309.17,516.00)(1.000,-1.000){2}{\rule{0.400pt}{0.241pt}}
\multiput(311.60,508.77)(0.468,-1.943){5}{\rule{0.113pt}{1.500pt}}
\multiput(310.17,511.89)(4.000,-10.887){2}{\rule{0.400pt}{0.750pt}}
\multiput(315.61,495.05)(0.447,-2.025){3}{\rule{0.108pt}{1.433pt}}
\multiput(314.17,498.03)(3.000,-7.025){2}{\rule{0.400pt}{0.717pt}}
\multiput(318.60,486.85)(0.468,-1.212){5}{\rule{0.113pt}{1.000pt}}
\multiput(317.17,488.92)(4.000,-6.924){2}{\rule{0.400pt}{0.500pt}}
\multiput(322.60,478.26)(0.468,-1.066){5}{\rule{0.113pt}{0.900pt}}
\multiput(321.17,480.13)(4.000,-6.132){2}{\rule{0.400pt}{0.450pt}}
\multiput(326.60,471.09)(0.468,-0.774){5}{\rule{0.113pt}{0.700pt}}
\multiput(325.17,472.55)(4.000,-4.547){2}{\rule{0.400pt}{0.350pt}}
\multiput(330.60,465.51)(0.468,-0.627){5}{\rule{0.113pt}{0.600pt}}
\multiput(329.17,466.75)(4.000,-3.755){2}{\rule{0.400pt}{0.300pt}}
\multiput(334.60,460.51)(0.468,-0.627){5}{\rule{0.113pt}{0.600pt}}
\multiput(333.17,461.75)(4.000,-3.755){2}{\rule{0.400pt}{0.300pt}}
\multiput(338.00,456.95)(0.685,-0.447){3}{\rule{0.633pt}{0.108pt}}
\multiput(338.00,457.17)(2.685,-3.000){2}{\rule{0.317pt}{0.400pt}}
\multiput(342.00,453.95)(0.685,-0.447){3}{\rule{0.633pt}{0.108pt}}
\multiput(342.00,454.17)(2.685,-3.000){2}{\rule{0.317pt}{0.400pt}}
\multiput(346.00,450.95)(0.685,-0.447){3}{\rule{0.633pt}{0.108pt}}
\multiput(346.00,451.17)(2.685,-3.000){2}{\rule{0.317pt}{0.400pt}}
\put(350,447.17){\rule{0.900pt}{0.400pt}}
\multiput(350.00,448.17)(2.132,-2.000){2}{\rule{0.450pt}{0.400pt}}
\put(354,445.67){\rule{0.964pt}{0.400pt}}
\multiput(354.00,446.17)(2.000,-1.000){2}{\rule{0.482pt}{0.400pt}}
\put(358,444.17){\rule{0.900pt}{0.400pt}}
\multiput(358.00,445.17)(2.132,-2.000){2}{\rule{0.450pt}{0.400pt}}
\put(362,442.67){\rule{0.964pt}{0.400pt}}
\multiput(362.00,443.17)(2.000,-1.000){2}{\rule{0.482pt}{0.400pt}}
\put(366,441.67){\rule{0.964pt}{0.400pt}}
\multiput(366.00,442.17)(2.000,-1.000){2}{\rule{0.482pt}{0.400pt}}
\put(370,440.67){\rule{0.964pt}{0.400pt}}
\multiput(370.00,441.17)(2.000,-1.000){2}{\rule{0.482pt}{0.400pt}}
\put(374,439.67){\rule{0.964pt}{0.400pt}}
\multiput(374.00,440.17)(2.000,-1.000){2}{\rule{0.482pt}{0.400pt}}
\put(382,438.67){\rule{0.723pt}{0.400pt}}
\multiput(382.00,439.17)(1.500,-1.000){2}{\rule{0.361pt}{0.400pt}}
\put(378.0,440.0){\rule[-0.200pt]{0.964pt}{0.400pt}}
\put(389,437.67){\rule{0.964pt}{0.400pt}}
\multiput(389.00,438.17)(2.000,-1.000){2}{\rule{0.482pt}{0.400pt}}
\put(385.0,439.0){\rule[-0.200pt]{0.964pt}{0.400pt}}
\put(401,436.67){\rule{0.964pt}{0.400pt}}
\multiput(401.00,437.17)(2.000,-1.000){2}{\rule{0.482pt}{0.400pt}}
\put(393.0,438.0){\rule[-0.200pt]{1.927pt}{0.400pt}}
\put(441,435.67){\rule{0.964pt}{0.400pt}}
\multiput(441.00,436.17)(2.000,-1.000){2}{\rule{0.482pt}{0.400pt}}
\put(405.0,437.0){\rule[-0.200pt]{8.672pt}{0.400pt}}
\put(445.0,436.0){\rule[-0.200pt]{39.748pt}{0.400pt}}
\put(269,213){\makebox(0,0){$\triangle$}}
\put(366,443){\makebox(0,0){$\triangle$}}
\put(464,435){\makebox(0,0){$\triangle$}}
\put(561,430){\makebox(0,0){$\triangle$}}
\put(309.67,514){\rule{0.400pt}{0.723pt}}
\multiput(309.17,515.50)(1.000,-1.500){2}{\rule{0.400pt}{0.361pt}}
\multiput(311.60,508.60)(0.468,-1.651){5}{\rule{0.113pt}{1.300pt}}
\multiput(310.17,511.30)(4.000,-9.302){2}{\rule{0.400pt}{0.650pt}}
\multiput(315.61,495.50)(0.447,-2.248){3}{\rule{0.108pt}{1.567pt}}
\multiput(314.17,498.75)(3.000,-7.748){2}{\rule{0.400pt}{0.783pt}}
\multiput(318.60,486.85)(0.468,-1.212){5}{\rule{0.113pt}{1.000pt}}
\multiput(317.17,488.92)(4.000,-6.924){2}{\rule{0.400pt}{0.500pt}}
\multiput(322.60,478.68)(0.468,-0.920){5}{\rule{0.113pt}{0.800pt}}
\multiput(321.17,480.34)(4.000,-5.340){2}{\rule{0.400pt}{0.400pt}}
\multiput(326.60,471.68)(0.468,-0.920){5}{\rule{0.113pt}{0.800pt}}
\multiput(325.17,473.34)(4.000,-5.340){2}{\rule{0.400pt}{0.400pt}}
\multiput(330.60,465.51)(0.468,-0.627){5}{\rule{0.113pt}{0.600pt}}
\multiput(329.17,466.75)(4.000,-3.755){2}{\rule{0.400pt}{0.300pt}}
\multiput(334.00,461.94)(0.481,-0.468){5}{\rule{0.500pt}{0.113pt}}
\multiput(334.00,462.17)(2.962,-4.000){2}{\rule{0.250pt}{0.400pt}}
\multiput(338.00,457.94)(0.481,-0.468){5}{\rule{0.500pt}{0.113pt}}
\multiput(338.00,458.17)(2.962,-4.000){2}{\rule{0.250pt}{0.400pt}}
\multiput(342.00,453.95)(0.685,-0.447){3}{\rule{0.633pt}{0.108pt}}
\multiput(342.00,454.17)(2.685,-3.000){2}{\rule{0.317pt}{0.400pt}}
\put(346,450.17){\rule{0.900pt}{0.400pt}}
\multiput(346.00,451.17)(2.132,-2.000){2}{\rule{0.450pt}{0.400pt}}
\put(350,448.17){\rule{0.900pt}{0.400pt}}
\multiput(350.00,449.17)(2.132,-2.000){2}{\rule{0.450pt}{0.400pt}}
\put(354,446.17){\rule{0.900pt}{0.400pt}}
\multiput(354.00,447.17)(2.132,-2.000){2}{\rule{0.450pt}{0.400pt}}
\put(358,444.17){\rule{0.900pt}{0.400pt}}
\multiput(358.00,445.17)(2.132,-2.000){2}{\rule{0.450pt}{0.400pt}}
\put(362,442.67){\rule{0.964pt}{0.400pt}}
\multiput(362.00,443.17)(2.000,-1.000){2}{\rule{0.482pt}{0.400pt}}
\put(366,441.67){\rule{0.964pt}{0.400pt}}
\multiput(366.00,442.17)(2.000,-1.000){2}{\rule{0.482pt}{0.400pt}}
\put(370,440.67){\rule{0.964pt}{0.400pt}}
\multiput(370.00,441.17)(2.000,-1.000){2}{\rule{0.482pt}{0.400pt}}
\put(374,439.67){\rule{0.964pt}{0.400pt}}
\multiput(374.00,440.17)(2.000,-1.000){2}{\rule{0.482pt}{0.400pt}}
\put(382,438.67){\rule{0.723pt}{0.400pt}}
\multiput(382.00,439.17)(1.500,-1.000){2}{\rule{0.361pt}{0.400pt}}
\put(378.0,440.0){\rule[-0.200pt]{0.964pt}{0.400pt}}
\put(389,437.67){\rule{0.964pt}{0.400pt}}
\multiput(389.00,438.17)(2.000,-1.000){2}{\rule{0.482pt}{0.400pt}}
\put(385.0,439.0){\rule[-0.200pt]{0.964pt}{0.400pt}}
\put(405,436.67){\rule{0.964pt}{0.400pt}}
\multiput(405.00,437.17)(2.000,-1.000){2}{\rule{0.482pt}{0.400pt}}
\put(393.0,438.0){\rule[-0.200pt]{2.891pt}{0.400pt}}
\put(441,435.67){\rule{0.964pt}{0.400pt}}
\multiput(441.00,436.17)(2.000,-1.000){2}{\rule{0.482pt}{0.400pt}}
\put(409.0,437.0){\rule[-0.200pt]{7.709pt}{0.400pt}}
\put(445.0,436.0){\rule[-0.200pt]{39.748pt}{0.400pt}}
\end{picture}

%% file: fig13.tex
\setlength{\unitlength}{0.240900pt}
\ifx\plotpoint\undefined\newsavebox{\plotpoint}\fi
\sbox{\plotpoint}{\rule[-0.200pt]{0.400pt}{0.400pt}}%
\begin{picture}(580,540)(0,0)
\font\gnuplot=cmr10 at 10pt
\gnuplot
\sbox{\plotpoint}{\rule[-0.200pt]{0.400pt}{0.400pt}}%
\put(220.0,315.0){\rule[-0.200pt]{93.951pt}{0.400pt}}
\put(220.0,113.0){\rule[-0.200pt]{4.818pt}{0.400pt}}
\put(198,113){\makebox(0,0)[r]{-0.01}}
\put(590.0,113.0){\rule[-0.200pt]{4.818pt}{0.400pt}}
\put(220.0,214.0){\rule[-0.200pt]{4.818pt}{0.400pt}}
\put(198,214){\makebox(0,0)[r]{-0.005}}
\put(590.0,214.0){\rule[-0.200pt]{4.818pt}{0.400pt}}
\put(220.0,315.0){\rule[-0.200pt]{4.818pt}{0.400pt}}
\put(198,315){\makebox(0,0)[r]{0}}
\put(590.0,315.0){\rule[-0.200pt]{4.818pt}{0.400pt}}
\put(220.0,416.0){\rule[-0.200pt]{4.818pt}{0.400pt}}
\put(198,416){\makebox(0,0)[r]{0.005}}
\put(590.0,416.0){\rule[-0.200pt]{4.818pt}{0.400pt}}
\put(220.0,517.0){\rule[-0.200pt]{4.818pt}{0.400pt}}
\put(198,517){\makebox(0,0)[r]{0.01}}
\put(590.0,517.0){\rule[-0.200pt]{4.818pt}{0.400pt}}
\put(269.0,113.0){\rule[-0.200pt]{0.400pt}{4.818pt}}
\put(269,68){\makebox(0,0){1}}
\put(269.0,497.0){\rule[-0.200pt]{0.400pt}{4.818pt}}
\put(366.0,113.0){\rule[-0.200pt]{0.400pt}{4.818pt}}
\put(366,68){\makebox(0,0){2}}
\put(366.0,497.0){\rule[-0.200pt]{0.400pt}{4.818pt}}
\put(464.0,113.0){\rule[-0.200pt]{0.400pt}{4.818pt}}
\put(464,68){\makebox(0,0){3}}
\put(464.0,497.0){\rule[-0.200pt]{0.400pt}{4.818pt}}
\put(561.0,113.0){\rule[-0.200pt]{0.400pt}{4.818pt}}
\put(561,68){\makebox(0,0){4}}
\put(561.0,497.0){\rule[-0.200pt]{0.400pt}{4.818pt}}
\put(220.0,113.0){\rule[-0.200pt]{93.951pt}{0.400pt}}
\put(610.0,113.0){\rule[-0.200pt]{0.400pt}{97.324pt}}
\put(220.0,517.0){\rule[-0.200pt]{93.951pt}{0.400pt}}
\put(65,315){\makebox(0,0){$G_R$}}
\put(415,23){\makebox(0,0){$d$}}
\put(220.0,113.0){\rule[-0.200pt]{0.400pt}{97.324pt}}
\put(366,395){\raisebox{-.8pt}{\makebox(0,0){$\Diamond$}}}
\put(464,303){\raisebox{-.8pt}{\makebox(0,0){$\Diamond$}}}
\put(561,325){\raisebox{-.8pt}{\makebox(0,0){$\Diamond$}}}
\put(366.0,375.0){\rule[-0.200pt]{0.400pt}{9.636pt}}
\put(356.0,375.0){\rule[-0.200pt]{4.818pt}{0.400pt}}
\put(356.0,415.0){\rule[-0.200pt]{4.818pt}{0.400pt}}
\put(464.0,279.0){\rule[-0.200pt]{0.400pt}{11.322pt}}
\put(454.0,279.0){\rule[-0.200pt]{4.818pt}{0.400pt}}
\put(454.0,326.0){\rule[-0.200pt]{4.818pt}{0.400pt}}
\put(561.0,292.0){\rule[-0.200pt]{0.400pt}{15.899pt}}
\put(551.0,292.0){\rule[-0.200pt]{4.818pt}{0.400pt}}
\put(551.0,358.0){\rule[-0.200pt]{4.818pt}{0.400pt}}
\multiput(343.61,504.96)(0.447,-4.481){3}{\rule{0.108pt}{2.900pt}}
\multiput(342.17,510.98)(3.000,-14.981){2}{\rule{0.400pt}{1.450pt}}
\multiput(346.60,483.96)(0.468,-3.990){5}{\rule{0.113pt}{2.900pt}}
\multiput(345.17,489.98)(4.000,-21.981){2}{\rule{0.400pt}{1.450pt}}
\multiput(350.60,458.04)(0.468,-3.259){5}{\rule{0.113pt}{2.400pt}}
\multiput(349.17,463.02)(4.000,-18.019){2}{\rule{0.400pt}{1.200pt}}
\multiput(354.60,436.70)(0.468,-2.674){5}{\rule{0.113pt}{2.000pt}}
\multiput(353.17,440.85)(4.000,-14.849){2}{\rule{0.400pt}{1.000pt}}
\multiput(358.60,418.53)(0.468,-2.382){5}{\rule{0.113pt}{1.800pt}}
\multiput(357.17,422.26)(4.000,-13.264){2}{\rule{0.400pt}{0.900pt}}
\multiput(362.60,402.77)(0.468,-1.943){5}{\rule{0.113pt}{1.500pt}}
\multiput(361.17,405.89)(4.000,-10.887){2}{\rule{0.400pt}{0.750pt}}
\multiput(366.60,389.60)(0.468,-1.651){5}{\rule{0.113pt}{1.300pt}}
\multiput(365.17,392.30)(4.000,-9.302){2}{\rule{0.400pt}{0.650pt}}
\multiput(370.60,378.43)(0.468,-1.358){5}{\rule{0.113pt}{1.100pt}}
\multiput(369.17,380.72)(4.000,-7.717){2}{\rule{0.400pt}{0.550pt}}
\multiput(374.60,369.26)(0.468,-1.066){5}{\rule{0.113pt}{0.900pt}}
\multiput(373.17,371.13)(4.000,-6.132){2}{\rule{0.400pt}{0.450pt}}
\multiput(378.60,361.68)(0.468,-0.920){5}{\rule{0.113pt}{0.800pt}}
\multiput(377.17,363.34)(4.000,-5.340){2}{\rule{0.400pt}{0.400pt}}
\multiput(382.61,353.71)(0.447,-1.355){3}{\rule{0.108pt}{1.033pt}}
\multiput(381.17,355.86)(3.000,-4.855){2}{\rule{0.400pt}{0.517pt}}
\multiput(385.60,348.51)(0.468,-0.627){5}{\rule{0.113pt}{0.600pt}}
\multiput(384.17,349.75)(4.000,-3.755){2}{\rule{0.400pt}{0.300pt}}
\multiput(389.00,344.94)(0.481,-0.468){5}{\rule{0.500pt}{0.113pt}}
\multiput(389.00,345.17)(2.962,-4.000){2}{\rule{0.250pt}{0.400pt}}
\multiput(393.00,340.94)(0.481,-0.468){5}{\rule{0.500pt}{0.113pt}}
\multiput(393.00,341.17)(2.962,-4.000){2}{\rule{0.250pt}{0.400pt}}
\multiput(397.00,336.95)(0.685,-0.447){3}{\rule{0.633pt}{0.108pt}}
\multiput(397.00,337.17)(2.685,-3.000){2}{\rule{0.317pt}{0.400pt}}
\multiput(401.00,333.95)(0.685,-0.447){3}{\rule{0.633pt}{0.108pt}}
\multiput(401.00,334.17)(2.685,-3.000){2}{\rule{0.317pt}{0.400pt}}
\multiput(405.00,330.95)(0.685,-0.447){3}{\rule{0.633pt}{0.108pt}}
\multiput(405.00,331.17)(2.685,-3.000){2}{\rule{0.317pt}{0.400pt}}
\put(409,327.17){\rule{0.900pt}{0.400pt}}
\multiput(409.00,328.17)(2.132,-2.000){2}{\rule{0.450pt}{0.400pt}}
\put(413,325.67){\rule{0.964pt}{0.400pt}}
\multiput(413.00,326.17)(2.000,-1.000){2}{\rule{0.482pt}{0.400pt}}
\put(417,324.17){\rule{0.900pt}{0.400pt}}
\multiput(417.00,325.17)(2.132,-2.000){2}{\rule{0.450pt}{0.400pt}}
\put(421,322.67){\rule{0.964pt}{0.400pt}}
\multiput(421.00,323.17)(2.000,-1.000){2}{\rule{0.482pt}{0.400pt}}
\put(425,321.67){\rule{0.964pt}{0.400pt}}
\multiput(425.00,322.17)(2.000,-1.000){2}{\rule{0.482pt}{0.400pt}}
\put(429,320.67){\rule{0.964pt}{0.400pt}}
\multiput(429.00,321.17)(2.000,-1.000){2}{\rule{0.482pt}{0.400pt}}
\put(433,319.67){\rule{0.964pt}{0.400pt}}
\multiput(433.00,320.17)(2.000,-1.000){2}{\rule{0.482pt}{0.400pt}}
\put(437,318.67){\rule{0.964pt}{0.400pt}}
\multiput(437.00,319.17)(2.000,-1.000){2}{\rule{0.482pt}{0.400pt}}
\put(445,317.67){\rule{0.723pt}{0.400pt}}
\multiput(445.00,318.17)(1.500,-1.000){2}{\rule{0.361pt}{0.400pt}}
\put(441.0,319.0){\rule[-0.200pt]{0.964pt}{0.400pt}}
\put(452,316.67){\rule{0.964pt}{0.400pt}}
\multiput(452.00,317.17)(2.000,-1.000){2}{\rule{0.482pt}{0.400pt}}
\put(448.0,318.0){\rule[-0.200pt]{0.964pt}{0.400pt}}
\put(468,315.67){\rule{0.964pt}{0.400pt}}
\multiput(468.00,316.17)(2.000,-1.000){2}{\rule{0.482pt}{0.400pt}}
\put(456.0,317.0){\rule[-0.200pt]{2.891pt}{0.400pt}}
\put(496,314.67){\rule{0.964pt}{0.400pt}}
\multiput(496.00,315.17)(2.000,-1.000){2}{\rule{0.482pt}{0.400pt}}
\put(472.0,316.0){\rule[-0.200pt]{5.782pt}{0.400pt}}
\put(500.0,315.0){\rule[-0.200pt]{26.499pt}{0.400pt}}
\put(366,399){\raisebox{-.8pt}{\makebox(0,0){$\Box$}}}
\put(464,288){\raisebox{-.8pt}{\makebox(0,0){$\Box$}}}
\put(561,307){\raisebox{-.8pt}{\makebox(0,0){$\Box$}}}
\put(366.0,373.0){\rule[-0.200pt]{0.400pt}{12.286pt}}
\put(356.0,373.0){\rule[-0.200pt]{4.818pt}{0.400pt}}
\put(356.0,424.0){\rule[-0.200pt]{4.818pt}{0.400pt}}
\put(464.0,265.0){\rule[-0.200pt]{0.400pt}{11.081pt}}
\put(454.0,265.0){\rule[-0.200pt]{4.818pt}{0.400pt}}
\put(454.0,311.0){\rule[-0.200pt]{4.818pt}{0.400pt}}
\put(561.0,281.0){\rule[-0.200pt]{0.400pt}{12.768pt}}
\put(551.0,281.0){\rule[-0.200pt]{4.818pt}{0.400pt}}
\put(551.0,334.0){\rule[-0.200pt]{4.818pt}{0.400pt}}
\put(344.67,512){\rule{0.400pt}{1.204pt}}
\multiput(344.17,514.50)(1.000,-2.500){2}{\rule{0.400pt}{0.602pt}}
\multiput(346.60,498.30)(0.468,-4.575){5}{\rule{0.113pt}{3.300pt}}
\multiput(345.17,505.15)(4.000,-25.151){2}{\rule{0.400pt}{1.650pt}}
\multiput(350.60,468.79)(0.468,-3.698){5}{\rule{0.113pt}{2.700pt}}
\multiput(349.17,474.40)(4.000,-20.396){2}{\rule{0.400pt}{1.350pt}}
\multiput(354.60,444.45)(0.468,-3.113){5}{\rule{0.113pt}{2.300pt}}
\multiput(353.17,449.23)(4.000,-17.226){2}{\rule{0.400pt}{1.150pt}}
\multiput(358.60,424.11)(0.468,-2.528){5}{\rule{0.113pt}{1.900pt}}
\multiput(357.17,428.06)(4.000,-14.056){2}{\rule{0.400pt}{0.950pt}}
\multiput(362.60,407.36)(0.468,-2.090){5}{\rule{0.113pt}{1.600pt}}
\multiput(361.17,410.68)(4.000,-11.679){2}{\rule{0.400pt}{0.800pt}}
\multiput(366.60,393.19)(0.468,-1.797){5}{\rule{0.113pt}{1.400pt}}
\multiput(365.17,396.09)(4.000,-10.094){2}{\rule{0.400pt}{0.700pt}}
\multiput(370.60,381.02)(0.468,-1.505){5}{\rule{0.113pt}{1.200pt}}
\multiput(369.17,383.51)(4.000,-8.509){2}{\rule{0.400pt}{0.600pt}}
\multiput(374.60,370.85)(0.468,-1.212){5}{\rule{0.113pt}{1.000pt}}
\multiput(373.17,372.92)(4.000,-6.924){2}{\rule{0.400pt}{0.500pt}}
\multiput(378.60,362.26)(0.468,-1.066){5}{\rule{0.113pt}{0.900pt}}
\multiput(377.17,364.13)(4.000,-6.132){2}{\rule{0.400pt}{0.450pt}}
\multiput(382.61,353.71)(0.447,-1.355){3}{\rule{0.108pt}{1.033pt}}
\multiput(381.17,355.86)(3.000,-4.855){2}{\rule{0.400pt}{0.517pt}}
\multiput(385.60,348.51)(0.468,-0.627){5}{\rule{0.113pt}{0.600pt}}
\multiput(384.17,349.75)(4.000,-3.755){2}{\rule{0.400pt}{0.300pt}}
\multiput(389.60,343.51)(0.468,-0.627){5}{\rule{0.113pt}{0.600pt}}
\multiput(388.17,344.75)(4.000,-3.755){2}{\rule{0.400pt}{0.300pt}}
\multiput(393.00,339.94)(0.481,-0.468){5}{\rule{0.500pt}{0.113pt}}
\multiput(393.00,340.17)(2.962,-4.000){2}{\rule{0.250pt}{0.400pt}}
\multiput(397.00,335.95)(0.685,-0.447){3}{\rule{0.633pt}{0.108pt}}
\multiput(397.00,336.17)(2.685,-3.000){2}{\rule{0.317pt}{0.400pt}}
\multiput(401.00,332.95)(0.685,-0.447){3}{\rule{0.633pt}{0.108pt}}
\multiput(401.00,333.17)(2.685,-3.000){2}{\rule{0.317pt}{0.400pt}}
\put(405,329.17){\rule{0.900pt}{0.400pt}}
\multiput(405.00,330.17)(2.132,-2.000){2}{\rule{0.450pt}{0.400pt}}
\put(409,327.17){\rule{0.900pt}{0.400pt}}
\multiput(409.00,328.17)(2.132,-2.000){2}{\rule{0.450pt}{0.400pt}}
\put(413,325.17){\rule{0.900pt}{0.400pt}}
\multiput(413.00,326.17)(2.132,-2.000){2}{\rule{0.450pt}{0.400pt}}
\put(417,323.17){\rule{0.900pt}{0.400pt}}
\multiput(417.00,324.17)(2.132,-2.000){2}{\rule{0.450pt}{0.400pt}}
\put(421,321.67){\rule{0.964pt}{0.400pt}}
\multiput(421.00,322.17)(2.000,-1.000){2}{\rule{0.482pt}{0.400pt}}
\put(425,320.67){\rule{0.964pt}{0.400pt}}
\multiput(425.00,321.17)(2.000,-1.000){2}{\rule{0.482pt}{0.400pt}}
\put(429,319.67){\rule{0.964pt}{0.400pt}}
\multiput(429.00,320.17)(2.000,-1.000){2}{\rule{0.482pt}{0.400pt}}
\put(433,318.67){\rule{0.964pt}{0.400pt}}
\multiput(433.00,319.17)(2.000,-1.000){2}{\rule{0.482pt}{0.400pt}}
\put(441,317.67){\rule{0.964pt}{0.400pt}}
\multiput(441.00,318.17)(2.000,-1.000){2}{\rule{0.482pt}{0.400pt}}
\put(437.0,319.0){\rule[-0.200pt]{0.964pt}{0.400pt}}
\put(448,316.67){\rule{0.964pt}{0.400pt}}
\multiput(448.00,317.17)(2.000,-1.000){2}{\rule{0.482pt}{0.400pt}}
\put(445.0,318.0){\rule[-0.200pt]{0.723pt}{0.400pt}}
\put(460,315.67){\rule{0.964pt}{0.400pt}}
\multiput(460.00,316.17)(2.000,-1.000){2}{\rule{0.482pt}{0.400pt}}
\put(452.0,317.0){\rule[-0.200pt]{1.927pt}{0.400pt}}
\put(492,314.67){\rule{0.964pt}{0.400pt}}
\multiput(492.00,315.17)(2.000,-1.000){2}{\rule{0.482pt}{0.400pt}}
\put(464.0,316.0){\rule[-0.200pt]{6.745pt}{0.400pt}}
\put(496.0,315.0){\rule[-0.200pt]{27.463pt}{0.400pt}}
\put(366,401){\makebox(0,0){$\triangle$}}
\put(464,296){\makebox(0,0){$\triangle$}}
\put(561,237){\makebox(0,0){$\triangle$}}
\put(366.0,383.0){\rule[-0.200pt]{0.400pt}{8.672pt}}
\put(356.0,383.0){\rule[-0.200pt]{4.818pt}{0.400pt}}
\put(356.0,419.0){\rule[-0.200pt]{4.818pt}{0.400pt}}
\put(464.0,272.0){\rule[-0.200pt]{0.400pt}{11.322pt}}
\put(454.0,272.0){\rule[-0.200pt]{4.818pt}{0.400pt}}
\put(454.0,319.0){\rule[-0.200pt]{4.818pt}{0.400pt}}
\put(561.0,205.0){\rule[-0.200pt]{0.400pt}{15.418pt}}
\put(551.0,205.0){\rule[-0.200pt]{4.818pt}{0.400pt}}
\put(551.0,269.0){\rule[-0.200pt]{4.818pt}{0.400pt}}
\put(346,517){\usebox{\plotpoint}}
\multiput(346.60,503.30)(0.468,-4.575){5}{\rule{0.113pt}{3.300pt}}
\multiput(345.17,510.15)(4.000,-25.151){2}{\rule{0.400pt}{1.650pt}}
\multiput(350.60,473.79)(0.468,-3.698){5}{\rule{0.113pt}{2.700pt}}
\multiput(349.17,479.40)(4.000,-20.396){2}{\rule{0.400pt}{1.350pt}}
\multiput(354.60,449.45)(0.468,-3.113){5}{\rule{0.113pt}{2.300pt}}
\multiput(353.17,454.23)(4.000,-17.226){2}{\rule{0.400pt}{1.150pt}}
\multiput(358.60,428.70)(0.468,-2.674){5}{\rule{0.113pt}{2.000pt}}
\multiput(357.17,432.85)(4.000,-14.849){2}{\rule{0.400pt}{1.000pt}}
\multiput(362.60,410.94)(0.468,-2.236){5}{\rule{0.113pt}{1.700pt}}
\multiput(361.17,414.47)(4.000,-12.472){2}{\rule{0.400pt}{0.850pt}}
\multiput(366.60,396.19)(0.468,-1.797){5}{\rule{0.113pt}{1.400pt}}
\multiput(365.17,399.09)(4.000,-10.094){2}{\rule{0.400pt}{0.700pt}}
\multiput(370.60,384.02)(0.468,-1.505){5}{\rule{0.113pt}{1.200pt}}
\multiput(369.17,386.51)(4.000,-8.509){2}{\rule{0.400pt}{0.600pt}}
\multiput(374.60,373.43)(0.468,-1.358){5}{\rule{0.113pt}{1.100pt}}
\multiput(373.17,375.72)(4.000,-7.717){2}{\rule{0.400pt}{0.550pt}}
\multiput(378.60,364.26)(0.468,-1.066){5}{\rule{0.113pt}{0.900pt}}
\multiput(377.17,366.13)(4.000,-6.132){2}{\rule{0.400pt}{0.450pt}}
\multiput(382.61,355.71)(0.447,-1.355){3}{\rule{0.108pt}{1.033pt}}
\multiput(381.17,357.86)(3.000,-4.855){2}{\rule{0.400pt}{0.517pt}}
\multiput(385.60,350.51)(0.468,-0.627){5}{\rule{0.113pt}{0.600pt}}
\multiput(384.17,351.75)(4.000,-3.755){2}{\rule{0.400pt}{0.300pt}}
\multiput(389.60,345.51)(0.468,-0.627){5}{\rule{0.113pt}{0.600pt}}
\multiput(388.17,346.75)(4.000,-3.755){2}{\rule{0.400pt}{0.300pt}}
\multiput(393.00,341.94)(0.481,-0.468){5}{\rule{0.500pt}{0.113pt}}
\multiput(393.00,342.17)(2.962,-4.000){2}{\rule{0.250pt}{0.400pt}}
\multiput(397.00,337.94)(0.481,-0.468){5}{\rule{0.500pt}{0.113pt}}
\multiput(397.00,338.17)(2.962,-4.000){2}{\rule{0.250pt}{0.400pt}}
\multiput(401.00,333.95)(0.685,-0.447){3}{\rule{0.633pt}{0.108pt}}
\multiput(401.00,334.17)(2.685,-3.000){2}{\rule{0.317pt}{0.400pt}}
\put(405,330.17){\rule{0.900pt}{0.400pt}}
\multiput(405.00,331.17)(2.132,-2.000){2}{\rule{0.450pt}{0.400pt}}
\put(409,328.17){\rule{0.900pt}{0.400pt}}
\multiput(409.00,329.17)(2.132,-2.000){2}{\rule{0.450pt}{0.400pt}}
\put(413,326.17){\rule{0.900pt}{0.400pt}}
\multiput(413.00,327.17)(2.132,-2.000){2}{\rule{0.450pt}{0.400pt}}
\put(417,324.17){\rule{0.900pt}{0.400pt}}
\multiput(417.00,325.17)(2.132,-2.000){2}{\rule{0.450pt}{0.400pt}}
\put(421,322.67){\rule{0.964pt}{0.400pt}}
\multiput(421.00,323.17)(2.000,-1.000){2}{\rule{0.482pt}{0.400pt}}
\put(425,321.67){\rule{0.964pt}{0.400pt}}
\multiput(425.00,322.17)(2.000,-1.000){2}{\rule{0.482pt}{0.400pt}}
\put(429,320.67){\rule{0.964pt}{0.400pt}}
\multiput(429.00,321.17)(2.000,-1.000){2}{\rule{0.482pt}{0.400pt}}
\put(433,319.67){\rule{0.964pt}{0.400pt}}
\multiput(433.00,320.17)(2.000,-1.000){2}{\rule{0.482pt}{0.400pt}}
\put(437,318.67){\rule{0.964pt}{0.400pt}}
\multiput(437.00,319.17)(2.000,-1.000){2}{\rule{0.482pt}{0.400pt}}
\put(445,317.67){\rule{0.723pt}{0.400pt}}
\multiput(445.00,318.17)(1.500,-1.000){2}{\rule{0.361pt}{0.400pt}}
\put(441.0,319.0){\rule[-0.200pt]{0.964pt}{0.400pt}}
\put(452,316.67){\rule{0.964pt}{0.400pt}}
\multiput(452.00,317.17)(2.000,-1.000){2}{\rule{0.482pt}{0.400pt}}
\put(448.0,318.0){\rule[-0.200pt]{0.964pt}{0.400pt}}
\put(464,315.67){\rule{0.964pt}{0.400pt}}
\multiput(464.00,316.17)(2.000,-1.000){2}{\rule{0.482pt}{0.400pt}}
\put(456.0,317.0){\rule[-0.200pt]{1.927pt}{0.400pt}}
\put(496,314.67){\rule{0.964pt}{0.400pt}}
\multiput(496.00,315.17)(2.000,-1.000){2}{\rule{0.482pt}{0.400pt}}
\put(468.0,316.0){\rule[-0.200pt]{6.745pt}{0.400pt}}
\put(500.0,315.0){\rule[-0.200pt]{26.499pt}{0.400pt}}
\end{picture}

%% file: fig21.tex
\setlength{\unitlength}{0.240900pt}
\ifx\plotpoint\undefined\newsavebox{\plotpoint}\fi
\sbox{\plotpoint}{\rule[-0.200pt]{0.400pt}{0.400pt}}%
\begin{picture}(600,540)(0,0)
\font\gnuplot=cmr10 at 10pt
\gnuplot
\sbox{\plotpoint}{\rule[-0.200pt]{0.400pt}{0.400pt}}%
\put(220.0,171.0){\rule[-0.200pt]{93.951pt}{0.400pt}}
\put(220.0,113.0){\rule[-0.200pt]{4.818pt}{0.400pt}}
\put(198,113){\makebox(0,0)[r]{-0.001}}
\put(590.0,113.0){\rule[-0.200pt]{4.818pt}{0.400pt}}
\put(220.0,171.0){\rule[-0.200pt]{4.818pt}{0.400pt}}
\put(198,171){\makebox(0,0)[r]{0}}
\put(590.0,171.0){\rule[-0.200pt]{4.818pt}{0.400pt}}
\put(220.0,228.0){\rule[-0.200pt]{4.818pt}{0.400pt}}
\put(198,228){\makebox(0,0)[r]{0.001}}
\put(590.0,228.0){\rule[-0.200pt]{4.818pt}{0.400pt}}
\put(220.0,286.0){\rule[-0.200pt]{4.818pt}{0.400pt}}
\put(198,286){\makebox(0,0)[r]{0.002}}
\put(590.0,286.0){\rule[-0.200pt]{4.818pt}{0.400pt}}
\put(220.0,344.0){\rule[-0.200pt]{4.818pt}{0.400pt}}
\put(198,344){\makebox(0,0)[r]{0.003}}
\put(590.0,344.0){\rule[-0.200pt]{4.818pt}{0.400pt}}
\put(220.0,402.0){\rule[-0.200pt]{4.818pt}{0.400pt}}
\put(198,402){\makebox(0,0)[r]{0.004}}
\put(590.0,402.0){\rule[-0.200pt]{4.818pt}{0.400pt}}
\put(220.0,459.0){\rule[-0.200pt]{4.818pt}{0.400pt}}
\put(198,459){\makebox(0,0)[r]{0.005}}
\put(590.0,459.0){\rule[-0.200pt]{4.818pt}{0.400pt}}
\put(220.0,517.0){\rule[-0.200pt]{4.818pt}{0.400pt}}
\put(198,517){\makebox(0,0)[r]{0.006}}
\put(590.0,517.0){\rule[-0.200pt]{4.818pt}{0.400pt}}
\put(269.0,113.0){\rule[-0.200pt]{0.400pt}{4.818pt}}
\put(269,68){\makebox(0,0){1}}
\put(269.0,497.0){\rule[-0.200pt]{0.400pt}{4.818pt}}
\put(366.0,113.0){\rule[-0.200pt]{0.400pt}{4.818pt}}
\put(366,68){\makebox(0,0){2}}
\put(366.0,497.0){\rule[-0.200pt]{0.400pt}{4.818pt}}
\put(464.0,113.0){\rule[-0.200pt]{0.400pt}{4.818pt}}
\put(464,68){\makebox(0,0){3}}
\put(464.0,497.0){\rule[-0.200pt]{0.400pt}{4.818pt}}
\put(561.0,113.0){\rule[-0.200pt]{0.400pt}{4.818pt}}
\put(561,68){\makebox(0,0){4}}
\put(561.0,497.0){\rule[-0.200pt]{0.400pt}{4.818pt}}
\put(220.0,113.0){\rule[-0.200pt]{93.951pt}{0.400pt}}
\put(610.0,113.0){\rule[-0.200pt]{0.400pt}{97.324pt}}
\put(220.0,517.0){\rule[-0.200pt]{93.951pt}{0.400pt}}
\put(45,315){\makebox(0,0){$G_V$}}
\put(415,23){\makebox(0,0){$d$}}
\put(220.0,113.0){\rule[-0.200pt]{0.400pt}{97.324pt}}
\put(269,478){\raisebox{-.8pt}{\makebox(0,0){$\Diamond$}}}
\put(366,157){\raisebox{-.8pt}{\makebox(0,0){$\Diamond$}}}
\put(464,166){\raisebox{-.8pt}{\makebox(0,0){$\Diamond$}}}
\put(561,168){\raisebox{-.8pt}{\makebox(0,0){$\Diamond$}}}
\put(269,485){\raisebox{-.8pt}{\makebox(0,0){$\Box$}}}
\put(366,165){\raisebox{-.8pt}{\makebox(0,0){$\Box$}}}
\put(464,173){\raisebox{-.8pt}{\makebox(0,0){$\Box$}}}
\put(561,169){\raisebox{-.8pt}{\makebox(0,0){$\Box$}}}
\put(269,473){\makebox(0,0){$\triangle$}}
\put(366,156){\makebox(0,0){$\triangle$}}
\put(464,168){\makebox(0,0){$\triangle$}}
\put(561,167){\makebox(0,0){$\triangle$}}
\end{picture}

%% file: fig22.tex
\setlength{\unitlength}{0.240900pt}
\ifx\plotpoint\undefined\newsavebox{\plotpoint}\fi
\sbox{\plotpoint}{\rule[-0.200pt]{0.400pt}{0.400pt}}%
\begin{picture}(580,540)(0,0)
\font\gnuplot=cmr10 at 10pt
\gnuplot
\sbox{\plotpoint}{\rule[-0.200pt]{0.400pt}{0.400pt}}%
\put(220.0,436.0){\rule[-0.200pt]{93.951pt}{0.400pt}}
\put(220.0,113.0){\rule[-0.200pt]{4.818pt}{0.400pt}}
\put(198,113){\makebox(0,0)[r]{-2}}
\put(590.0,113.0){\rule[-0.200pt]{4.818pt}{0.400pt}}
\put(220.0,194.0){\rule[-0.200pt]{4.818pt}{0.400pt}}
\put(198,194){\makebox(0,0)[r]{-1.5}}
\put(590.0,194.0){\rule[-0.200pt]{4.818pt}{0.400pt}}
\put(220.0,275.0){\rule[-0.200pt]{4.818pt}{0.400pt}}
\put(198,275){\makebox(0,0)[r]{-1}}
\put(590.0,275.0){\rule[-0.200pt]{4.818pt}{0.400pt}}
\put(220.0,355.0){\rule[-0.200pt]{4.818pt}{0.400pt}}
\put(198,355){\makebox(0,0)[r]{-0.5}}
\put(590.0,355.0){\rule[-0.200pt]{4.818pt}{0.400pt}}
\put(220.0,436.0){\rule[-0.200pt]{4.818pt}{0.400pt}}
\put(198,436){\makebox(0,0)[r]{0}}
\put(590.0,436.0){\rule[-0.200pt]{4.818pt}{0.400pt}}
\put(220.0,517.0){\rule[-0.200pt]{4.818pt}{0.400pt}}
\put(198,517){\makebox(0,0)[r]{0.5}}
\put(590.0,517.0){\rule[-0.200pt]{4.818pt}{0.400pt}}
\put(269.0,113.0){\rule[-0.200pt]{0.400pt}{4.818pt}}
\put(269,68){\makebox(0,0){1}}
\put(269.0,497.0){\rule[-0.200pt]{0.400pt}{4.818pt}}
\put(366.0,113.0){\rule[-0.200pt]{0.400pt}{4.818pt}}
\put(366,68){\makebox(0,0){2}}
\put(366.0,497.0){\rule[-0.200pt]{0.400pt}{4.818pt}}
\put(464.0,113.0){\rule[-0.200pt]{0.400pt}{4.818pt}}
\put(464,68){\makebox(0,0){3}}
\put(464.0,497.0){\rule[-0.200pt]{0.400pt}{4.818pt}}
\put(561.0,113.0){\rule[-0.200pt]{0.400pt}{4.818pt}}
\put(561,68){\makebox(0,0){4}}
\put(561.0,497.0){\rule[-0.200pt]{0.400pt}{4.818pt}}
\put(220.0,113.0){\rule[-0.200pt]{93.951pt}{0.400pt}}
\put(610.0,113.0){\rule[-0.200pt]{0.400pt}{97.324pt}}
\put(220.0,517.0){\rule[-0.200pt]{93.951pt}{0.400pt}}
\put(65,315){\makebox(0,0){$G_R$}}
\put(415,23){\makebox(0,0){$d$}}
\put(220.0,113.0){\rule[-0.200pt]{0.400pt}{97.324pt}}
\put(269,173){\raisebox{-.8pt}{\makebox(0,0){$\Diamond$}}}
\put(366,455){\raisebox{-.8pt}{\makebox(0,0){$\Diamond$}}}
\put(464,440){\raisebox{-.8pt}{\makebox(0,0){$\Diamond$}}}
\put(561,431){\raisebox{-.8pt}{\makebox(0,0){$\Diamond$}}}
\put(305.17,513){\rule{0.400pt}{0.900pt}}
\multiput(304.17,515.13)(2.000,-2.132){2}{\rule{0.400pt}{0.450pt}}
\multiput(307.60,509.26)(0.468,-1.066){5}{\rule{0.113pt}{0.900pt}}
\multiput(306.17,511.13)(4.000,-6.132){2}{\rule{0.400pt}{0.450pt}}
\multiput(311.60,501.68)(0.468,-0.920){5}{\rule{0.113pt}{0.800pt}}
\multiput(310.17,503.34)(4.000,-5.340){2}{\rule{0.400pt}{0.400pt}}
\multiput(315.61,494.82)(0.447,-0.909){3}{\rule{0.108pt}{0.767pt}}
\multiput(314.17,496.41)(3.000,-3.409){2}{\rule{0.400pt}{0.383pt}}
\multiput(318.60,490.09)(0.468,-0.774){5}{\rule{0.113pt}{0.700pt}}
\multiput(317.17,491.55)(4.000,-4.547){2}{\rule{0.400pt}{0.350pt}}
\multiput(322.60,484.51)(0.468,-0.627){5}{\rule{0.113pt}{0.600pt}}
\multiput(321.17,485.75)(4.000,-3.755){2}{\rule{0.400pt}{0.300pt}}
\multiput(326.00,480.94)(0.481,-0.468){5}{\rule{0.500pt}{0.113pt}}
\multiput(326.00,481.17)(2.962,-4.000){2}{\rule{0.250pt}{0.400pt}}
\multiput(330.00,476.94)(0.481,-0.468){5}{\rule{0.500pt}{0.113pt}}
\multiput(330.00,477.17)(2.962,-4.000){2}{\rule{0.250pt}{0.400pt}}
\multiput(334.00,472.95)(0.685,-0.447){3}{\rule{0.633pt}{0.108pt}}
\multiput(334.00,473.17)(2.685,-3.000){2}{\rule{0.317pt}{0.400pt}}
\multiput(338.00,469.95)(0.685,-0.447){3}{\rule{0.633pt}{0.108pt}}
\multiput(338.00,470.17)(2.685,-3.000){2}{\rule{0.317pt}{0.400pt}}
\multiput(342.00,466.95)(0.685,-0.447){3}{\rule{0.633pt}{0.108pt}}
\multiput(342.00,467.17)(2.685,-3.000){2}{\rule{0.317pt}{0.400pt}}
\put(346,463.17){\rule{0.900pt}{0.400pt}}
\multiput(346.00,464.17)(2.132,-2.000){2}{\rule{0.450pt}{0.400pt}}
\multiput(350.00,461.95)(0.685,-0.447){3}{\rule{0.633pt}{0.108pt}}
\multiput(350.00,462.17)(2.685,-3.000){2}{\rule{0.317pt}{0.400pt}}
\put(354,458.17){\rule{0.900pt}{0.400pt}}
\multiput(354.00,459.17)(2.132,-2.000){2}{\rule{0.450pt}{0.400pt}}
\put(358,456.17){\rule{0.900pt}{0.400pt}}
\multiput(358.00,457.17)(2.132,-2.000){2}{\rule{0.450pt}{0.400pt}}
\put(362,454.67){\rule{0.964pt}{0.400pt}}
\multiput(362.00,455.17)(2.000,-1.000){2}{\rule{0.482pt}{0.400pt}}
\put(366,453.17){\rule{0.900pt}{0.400pt}}
\multiput(366.00,454.17)(2.132,-2.000){2}{\rule{0.450pt}{0.400pt}}
\put(370,451.67){\rule{0.964pt}{0.400pt}}
\multiput(370.00,452.17)(2.000,-1.000){2}{\rule{0.482pt}{0.400pt}}
\put(374,450.67){\rule{0.964pt}{0.400pt}}
\multiput(374.00,451.17)(2.000,-1.000){2}{\rule{0.482pt}{0.400pt}}
\put(378,449.17){\rule{0.900pt}{0.400pt}}
\multiput(378.00,450.17)(2.132,-2.000){2}{\rule{0.450pt}{0.400pt}}
\put(382,447.67){\rule{0.723pt}{0.400pt}}
\multiput(382.00,448.17)(1.500,-1.000){2}{\rule{0.361pt}{0.400pt}}
\put(385,446.67){\rule{0.964pt}{0.400pt}}
\multiput(385.00,447.17)(2.000,-1.000){2}{\rule{0.482pt}{0.400pt}}
\put(389,445.67){\rule{0.964pt}{0.400pt}}
\multiput(389.00,446.17)(2.000,-1.000){2}{\rule{0.482pt}{0.400pt}}
\put(397,444.67){\rule{0.964pt}{0.400pt}}
\multiput(397.00,445.17)(2.000,-1.000){2}{\rule{0.482pt}{0.400pt}}
\put(401,443.67){\rule{0.964pt}{0.400pt}}
\multiput(401.00,444.17)(2.000,-1.000){2}{\rule{0.482pt}{0.400pt}}
\put(393.0,446.0){\rule[-0.200pt]{0.964pt}{0.400pt}}
\put(409,442.67){\rule{0.964pt}{0.400pt}}
\multiput(409.00,443.17)(2.000,-1.000){2}{\rule{0.482pt}{0.400pt}}
\put(405.0,444.0){\rule[-0.200pt]{0.964pt}{0.400pt}}
\put(417,441.67){\rule{0.964pt}{0.400pt}}
\multiput(417.00,442.17)(2.000,-1.000){2}{\rule{0.482pt}{0.400pt}}
\put(413.0,443.0){\rule[-0.200pt]{0.964pt}{0.400pt}}
\put(425,440.67){\rule{0.964pt}{0.400pt}}
\multiput(425.00,441.17)(2.000,-1.000){2}{\rule{0.482pt}{0.400pt}}
\put(421.0,442.0){\rule[-0.200pt]{0.964pt}{0.400pt}}
\put(437,439.67){\rule{0.964pt}{0.400pt}}
\multiput(437.00,440.17)(2.000,-1.000){2}{\rule{0.482pt}{0.400pt}}
\put(429.0,441.0){\rule[-0.200pt]{1.927pt}{0.400pt}}
\put(448,438.67){\rule{0.964pt}{0.400pt}}
\multiput(448.00,439.17)(2.000,-1.000){2}{\rule{0.482pt}{0.400pt}}
\put(441.0,440.0){\rule[-0.200pt]{1.686pt}{0.400pt}}
\put(468,437.67){\rule{0.964pt}{0.400pt}}
\multiput(468.00,438.17)(2.000,-1.000){2}{\rule{0.482pt}{0.400pt}}
\put(452.0,439.0){\rule[-0.200pt]{3.854pt}{0.400pt}}
\put(500,436.67){\rule{0.964pt}{0.400pt}}
\multiput(500.00,437.17)(2.000,-1.000){2}{\rule{0.482pt}{0.400pt}}
\put(472.0,438.0){\rule[-0.200pt]{6.745pt}{0.400pt}}
\put(586,435.67){\rule{0.964pt}{0.400pt}}
\multiput(586.00,436.17)(2.000,-1.000){2}{\rule{0.482pt}{0.400pt}}
\put(504.0,437.0){\rule[-0.200pt]{19.754pt}{0.400pt}}
\put(590.0,436.0){\rule[-0.200pt]{4.818pt}{0.400pt}}
\put(269,154){\raisebox{-.8pt}{\makebox(0,0){$\Box$}}}
\put(366,457){\raisebox{-.8pt}{\makebox(0,0){$\Box$}}}
\put(464,441){\raisebox{-.8pt}{\makebox(0,0){$\Box$}}}
\put(561,430){\raisebox{-.8pt}{\makebox(0,0){$\Box$}}}
\multiput(303.60,513.68)(0.468,-0.920){5}{\rule{0.113pt}{0.800pt}}
\multiput(302.17,515.34)(4.000,-5.340){2}{\rule{0.400pt}{0.400pt}}
\multiput(307.60,506.68)(0.468,-0.920){5}{\rule{0.113pt}{0.800pt}}
\multiput(306.17,508.34)(4.000,-5.340){2}{\rule{0.400pt}{0.400pt}}
\multiput(311.60,500.09)(0.468,-0.774){5}{\rule{0.113pt}{0.700pt}}
\multiput(310.17,501.55)(4.000,-4.547){2}{\rule{0.400pt}{0.350pt}}
\multiput(315.61,493.82)(0.447,-0.909){3}{\rule{0.108pt}{0.767pt}}
\multiput(314.17,495.41)(3.000,-3.409){2}{\rule{0.400pt}{0.383pt}}
\multiput(318.60,489.51)(0.468,-0.627){5}{\rule{0.113pt}{0.600pt}}
\multiput(317.17,490.75)(4.000,-3.755){2}{\rule{0.400pt}{0.300pt}}
\multiput(322.00,485.94)(0.481,-0.468){5}{\rule{0.500pt}{0.113pt}}
\multiput(322.00,486.17)(2.962,-4.000){2}{\rule{0.250pt}{0.400pt}}
\multiput(326.00,481.94)(0.481,-0.468){5}{\rule{0.500pt}{0.113pt}}
\multiput(326.00,482.17)(2.962,-4.000){2}{\rule{0.250pt}{0.400pt}}
\multiput(330.00,477.95)(0.685,-0.447){3}{\rule{0.633pt}{0.108pt}}
\multiput(330.00,478.17)(2.685,-3.000){2}{\rule{0.317pt}{0.400pt}}
\multiput(334.00,474.94)(0.481,-0.468){5}{\rule{0.500pt}{0.113pt}}
\multiput(334.00,475.17)(2.962,-4.000){2}{\rule{0.250pt}{0.400pt}}
\put(338,470.17){\rule{0.900pt}{0.400pt}}
\multiput(338.00,471.17)(2.132,-2.000){2}{\rule{0.450pt}{0.400pt}}
\multiput(342.00,468.95)(0.685,-0.447){3}{\rule{0.633pt}{0.108pt}}
\multiput(342.00,469.17)(2.685,-3.000){2}{\rule{0.317pt}{0.400pt}}
\multiput(346.00,465.95)(0.685,-0.447){3}{\rule{0.633pt}{0.108pt}}
\multiput(346.00,466.17)(2.685,-3.000){2}{\rule{0.317pt}{0.400pt}}
\put(350,462.17){\rule{0.900pt}{0.400pt}}
\multiput(350.00,463.17)(2.132,-2.000){2}{\rule{0.450pt}{0.400pt}}
\put(354,460.17){\rule{0.900pt}{0.400pt}}
\multiput(354.00,461.17)(2.132,-2.000){2}{\rule{0.450pt}{0.400pt}}
\put(358,458.17){\rule{0.900pt}{0.400pt}}
\multiput(358.00,459.17)(2.132,-2.000){2}{\rule{0.450pt}{0.400pt}}
\put(362,456.67){\rule{0.964pt}{0.400pt}}
\multiput(362.00,457.17)(2.000,-1.000){2}{\rule{0.482pt}{0.400pt}}
\put(366,455.17){\rule{0.900pt}{0.400pt}}
\multiput(366.00,456.17)(2.132,-2.000){2}{\rule{0.450pt}{0.400pt}}
\put(370,453.67){\rule{0.964pt}{0.400pt}}
\multiput(370.00,454.17)(2.000,-1.000){2}{\rule{0.482pt}{0.400pt}}
\put(374,452.67){\rule{0.964pt}{0.400pt}}
\multiput(374.00,453.17)(2.000,-1.000){2}{\rule{0.482pt}{0.400pt}}
\put(378,451.17){\rule{0.900pt}{0.400pt}}
\multiput(378.00,452.17)(2.132,-2.000){2}{\rule{0.450pt}{0.400pt}}
\put(382,449.67){\rule{0.723pt}{0.400pt}}
\multiput(382.00,450.17)(1.500,-1.000){2}{\rule{0.361pt}{0.400pt}}
\put(385,448.67){\rule{0.964pt}{0.400pt}}
\multiput(385.00,449.17)(2.000,-1.000){2}{\rule{0.482pt}{0.400pt}}
\put(389,447.67){\rule{0.964pt}{0.400pt}}
\multiput(389.00,448.17)(2.000,-1.000){2}{\rule{0.482pt}{0.400pt}}
\put(397,446.67){\rule{0.964pt}{0.400pt}}
\multiput(397.00,447.17)(2.000,-1.000){2}{\rule{0.482pt}{0.400pt}}
\put(401,445.67){\rule{0.964pt}{0.400pt}}
\multiput(401.00,446.17)(2.000,-1.000){2}{\rule{0.482pt}{0.400pt}}
\put(405,444.67){\rule{0.964pt}{0.400pt}}
\multiput(405.00,445.17)(2.000,-1.000){2}{\rule{0.482pt}{0.400pt}}
\put(393.0,448.0){\rule[-0.200pt]{0.964pt}{0.400pt}}
\put(413,443.67){\rule{0.964pt}{0.400pt}}
\multiput(413.00,444.17)(2.000,-1.000){2}{\rule{0.482pt}{0.400pt}}
\put(409.0,445.0){\rule[-0.200pt]{0.964pt}{0.400pt}}
\put(421,442.67){\rule{0.964pt}{0.400pt}}
\multiput(421.00,443.17)(2.000,-1.000){2}{\rule{0.482pt}{0.400pt}}
\put(417.0,444.0){\rule[-0.200pt]{0.964pt}{0.400pt}}
\put(429,441.67){\rule{0.964pt}{0.400pt}}
\multiput(429.00,442.17)(2.000,-1.000){2}{\rule{0.482pt}{0.400pt}}
\put(425.0,443.0){\rule[-0.200pt]{0.964pt}{0.400pt}}
\put(441,440.67){\rule{0.964pt}{0.400pt}}
\multiput(441.00,441.17)(2.000,-1.000){2}{\rule{0.482pt}{0.400pt}}
\put(433.0,442.0){\rule[-0.200pt]{1.927pt}{0.400pt}}
\put(452,439.67){\rule{0.964pt}{0.400pt}}
\multiput(452.00,440.17)(2.000,-1.000){2}{\rule{0.482pt}{0.400pt}}
\put(445.0,441.0){\rule[-0.200pt]{1.686pt}{0.400pt}}
\put(468,438.67){\rule{0.964pt}{0.400pt}}
\multiput(468.00,439.17)(2.000,-1.000){2}{\rule{0.482pt}{0.400pt}}
\put(456.0,440.0){\rule[-0.200pt]{2.891pt}{0.400pt}}
\put(492,437.67){\rule{0.964pt}{0.400pt}}
\multiput(492.00,438.17)(2.000,-1.000){2}{\rule{0.482pt}{0.400pt}}
\put(472.0,439.0){\rule[-0.200pt]{4.818pt}{0.400pt}}
\put(527,436.67){\rule{0.964pt}{0.400pt}}
\multiput(527.00,437.17)(2.000,-1.000){2}{\rule{0.482pt}{0.400pt}}
\put(496.0,438.0){\rule[-0.200pt]{7.468pt}{0.400pt}}
\put(531.0,437.0){\rule[-0.200pt]{19.031pt}{0.400pt}}
\put(269,151){\makebox(0,0){$\triangle$}}
\put(366,452){\makebox(0,0){$\triangle$}}
\put(464,445){\makebox(0,0){$\triangle$}}
\put(561,428){\makebox(0,0){$\triangle$}}
\put(281.67,515){\rule{0.400pt}{0.482pt}}
\multiput(281.17,516.00)(1.000,-1.000){2}{\rule{0.400pt}{0.241pt}}
\multiput(283.60,511.68)(0.468,-0.920){5}{\rule{0.113pt}{0.800pt}}
\multiput(282.17,513.34)(4.000,-5.340){2}{\rule{0.400pt}{0.400pt}}
\multiput(287.60,505.09)(0.468,-0.774){5}{\rule{0.113pt}{0.700pt}}
\multiput(286.17,506.55)(4.000,-4.547){2}{\rule{0.400pt}{0.350pt}}
\multiput(291.60,499.51)(0.468,-0.627){5}{\rule{0.113pt}{0.600pt}}
\multiput(290.17,500.75)(4.000,-3.755){2}{\rule{0.400pt}{0.300pt}}
\multiput(295.60,494.51)(0.468,-0.627){5}{\rule{0.113pt}{0.600pt}}
\multiput(294.17,495.75)(4.000,-3.755){2}{\rule{0.400pt}{0.300pt}}
\multiput(299.00,490.94)(0.481,-0.468){5}{\rule{0.500pt}{0.113pt}}
\multiput(299.00,491.17)(2.962,-4.000){2}{\rule{0.250pt}{0.400pt}}
\multiput(303.00,486.94)(0.481,-0.468){5}{\rule{0.500pt}{0.113pt}}
\multiput(303.00,487.17)(2.962,-4.000){2}{\rule{0.250pt}{0.400pt}}
\multiput(307.00,482.94)(0.481,-0.468){5}{\rule{0.500pt}{0.113pt}}
\multiput(307.00,483.17)(2.962,-4.000){2}{\rule{0.250pt}{0.400pt}}
\multiput(311.00,478.95)(0.685,-0.447){3}{\rule{0.633pt}{0.108pt}}
\multiput(311.00,479.17)(2.685,-3.000){2}{\rule{0.317pt}{0.400pt}}
\multiput(315.00,475.95)(0.462,-0.447){3}{\rule{0.500pt}{0.108pt}}
\multiput(315.00,476.17)(1.962,-3.000){2}{\rule{0.250pt}{0.400pt}}
\multiput(318.00,472.95)(0.685,-0.447){3}{\rule{0.633pt}{0.108pt}}
\multiput(318.00,473.17)(2.685,-3.000){2}{\rule{0.317pt}{0.400pt}}
\put(322,469.17){\rule{0.900pt}{0.400pt}}
\multiput(322.00,470.17)(2.132,-2.000){2}{\rule{0.450pt}{0.400pt}}
\multiput(326.00,467.95)(0.685,-0.447){3}{\rule{0.633pt}{0.108pt}}
\multiput(326.00,468.17)(2.685,-3.000){2}{\rule{0.317pt}{0.400pt}}
\put(330,464.17){\rule{0.900pt}{0.400pt}}
\multiput(330.00,465.17)(2.132,-2.000){2}{\rule{0.450pt}{0.400pt}}
\put(334,462.17){\rule{0.900pt}{0.400pt}}
\multiput(334.00,463.17)(2.132,-2.000){2}{\rule{0.450pt}{0.400pt}}
\put(338,460.67){\rule{0.964pt}{0.400pt}}
\multiput(338.00,461.17)(2.000,-1.000){2}{\rule{0.482pt}{0.400pt}}
\put(342,459.17){\rule{0.900pt}{0.400pt}}
\multiput(342.00,460.17)(2.132,-2.000){2}{\rule{0.450pt}{0.400pt}}
\put(346,457.67){\rule{0.964pt}{0.400pt}}
\multiput(346.00,458.17)(2.000,-1.000){2}{\rule{0.482pt}{0.400pt}}
\put(350,456.17){\rule{0.900pt}{0.400pt}}
\multiput(350.00,457.17)(2.132,-2.000){2}{\rule{0.450pt}{0.400pt}}
\put(354,454.67){\rule{0.964pt}{0.400pt}}
\multiput(354.00,455.17)(2.000,-1.000){2}{\rule{0.482pt}{0.400pt}}
\put(358,453.67){\rule{0.964pt}{0.400pt}}
\multiput(358.00,454.17)(2.000,-1.000){2}{\rule{0.482pt}{0.400pt}}
\put(362,452.67){\rule{0.964pt}{0.400pt}}
\multiput(362.00,453.17)(2.000,-1.000){2}{\rule{0.482pt}{0.400pt}}
\put(366,451.67){\rule{0.964pt}{0.400pt}}
\multiput(366.00,452.17)(2.000,-1.000){2}{\rule{0.482pt}{0.400pt}}
\put(370,450.67){\rule{0.964pt}{0.400pt}}
\multiput(370.00,451.17)(2.000,-1.000){2}{\rule{0.482pt}{0.400pt}}
\put(374,449.67){\rule{0.964pt}{0.400pt}}
\multiput(374.00,450.17)(2.000,-1.000){2}{\rule{0.482pt}{0.400pt}}
\put(378,448.67){\rule{0.964pt}{0.400pt}}
\multiput(378.00,449.17)(2.000,-1.000){2}{\rule{0.482pt}{0.400pt}}
\put(385,447.67){\rule{0.964pt}{0.400pt}}
\multiput(385.00,448.17)(2.000,-1.000){2}{\rule{0.482pt}{0.400pt}}
\put(389,446.67){\rule{0.964pt}{0.400pt}}
\multiput(389.00,447.17)(2.000,-1.000){2}{\rule{0.482pt}{0.400pt}}
\put(382.0,449.0){\rule[-0.200pt]{0.723pt}{0.400pt}}
\put(397,445.67){\rule{0.964pt}{0.400pt}}
\multiput(397.00,446.17)(2.000,-1.000){2}{\rule{0.482pt}{0.400pt}}
\put(393.0,447.0){\rule[-0.200pt]{0.964pt}{0.400pt}}
\put(405,444.67){\rule{0.964pt}{0.400pt}}
\multiput(405.00,445.17)(2.000,-1.000){2}{\rule{0.482pt}{0.400pt}}
\put(401.0,446.0){\rule[-0.200pt]{0.964pt}{0.400pt}}
\put(413,443.67){\rule{0.964pt}{0.400pt}}
\multiput(413.00,444.17)(2.000,-1.000){2}{\rule{0.482pt}{0.400pt}}
\put(409.0,445.0){\rule[-0.200pt]{0.964pt}{0.400pt}}
\put(421,442.67){\rule{0.964pt}{0.400pt}}
\multiput(421.00,443.17)(2.000,-1.000){2}{\rule{0.482pt}{0.400pt}}
\put(417.0,444.0){\rule[-0.200pt]{0.964pt}{0.400pt}}
\put(433,441.67){\rule{0.964pt}{0.400pt}}
\multiput(433.00,442.17)(2.000,-1.000){2}{\rule{0.482pt}{0.400pt}}
\put(425.0,443.0){\rule[-0.200pt]{1.927pt}{0.400pt}}
\put(445,440.67){\rule{0.723pt}{0.400pt}}
\multiput(445.00,441.17)(1.500,-1.000){2}{\rule{0.361pt}{0.400pt}}
\put(437.0,442.0){\rule[-0.200pt]{1.927pt}{0.400pt}}
\put(464,439.67){\rule{0.964pt}{0.400pt}}
\multiput(464.00,440.17)(2.000,-1.000){2}{\rule{0.482pt}{0.400pt}}
\put(448.0,441.0){\rule[-0.200pt]{3.854pt}{0.400pt}}
\put(484,438.67){\rule{0.964pt}{0.400pt}}
\multiput(484.00,439.17)(2.000,-1.000){2}{\rule{0.482pt}{0.400pt}}
\put(468.0,440.0){\rule[-0.200pt]{3.854pt}{0.400pt}}
\put(515,437.67){\rule{0.964pt}{0.400pt}}
\multiput(515.00,438.17)(2.000,-1.000){2}{\rule{0.482pt}{0.400pt}}
\put(488.0,439.0){\rule[-0.200pt]{6.504pt}{0.400pt}}
\put(571,436.67){\rule{0.964pt}{0.400pt}}
\multiput(571.00,437.17)(2.000,-1.000){2}{\rule{0.482pt}{0.400pt}}
\put(519.0,438.0){\rule[-0.200pt]{12.527pt}{0.400pt}}
\put(575.0,437.0){\rule[-0.200pt]{8.431pt}{0.400pt}}
\end{picture}

%% file: fig23.tex
\setlength{\unitlength}{0.240900pt}
\ifx\plotpoint\undefined\newsavebox{\plotpoint}\fi
\sbox{\plotpoint}{\rule[-0.200pt]{0.400pt}{0.400pt}}%
\begin{picture}(580,540)(0,0)
\font\gnuplot=cmr10 at 10pt
\gnuplot
\sbox{\plotpoint}{\rule[-0.200pt]{0.400pt}{0.400pt}}%
\put(220.0,315.0){\rule[-0.200pt]{93.951pt}{0.400pt}}
\put(220.0,113.0){\rule[-0.200pt]{4.818pt}{0.400pt}}
\put(198,113){\makebox(0,0)[r]{-0.2}}
\put(590.0,113.0){\rule[-0.200pt]{4.818pt}{0.400pt}}
\put(220.0,164.0){\rule[-0.200pt]{4.818pt}{0.400pt}}
\put(198,164){\makebox(0,0)[r]{-0.15}}
\put(590.0,164.0){\rule[-0.200pt]{4.818pt}{0.400pt}}
\put(220.0,214.0){\rule[-0.200pt]{4.818pt}{0.400pt}}
\put(198,214){\makebox(0,0)[r]{-0.1}}
\put(590.0,214.0){\rule[-0.200pt]{4.818pt}{0.400pt}}
\put(220.0,265.0){\rule[-0.200pt]{4.818pt}{0.400pt}}
\put(198,265){\makebox(0,0)[r]{-0.05}}
\put(590.0,265.0){\rule[-0.200pt]{4.818pt}{0.400pt}}
\put(220.0,315.0){\rule[-0.200pt]{4.818pt}{0.400pt}}
\put(198,315){\makebox(0,0)[r]{0}}
\put(590.0,315.0){\rule[-0.200pt]{4.818pt}{0.400pt}}
\put(220.0,366.0){\rule[-0.200pt]{4.818pt}{0.400pt}}
\put(198,366){\makebox(0,0)[r]{0.05}}
\put(590.0,366.0){\rule[-0.200pt]{4.818pt}{0.400pt}}
\put(220.0,416.0){\rule[-0.200pt]{4.818pt}{0.400pt}}
\put(198,416){\makebox(0,0)[r]{0.1}}
\put(590.0,416.0){\rule[-0.200pt]{4.818pt}{0.400pt}}
\put(220.0,467.0){\rule[-0.200pt]{4.818pt}{0.400pt}}
\put(198,467){\makebox(0,0)[r]{0.15}}
\put(590.0,467.0){\rule[-0.200pt]{4.818pt}{0.400pt}}
\put(220.0,517.0){\rule[-0.200pt]{4.818pt}{0.400pt}}
\put(198,517){\makebox(0,0)[r]{0.2}}
\put(590.0,517.0){\rule[-0.200pt]{4.818pt}{0.400pt}}
\put(269.0,113.0){\rule[-0.200pt]{0.400pt}{4.818pt}}
\put(269,68){\makebox(0,0){1}}
\put(269.0,497.0){\rule[-0.200pt]{0.400pt}{4.818pt}}
\put(366.0,113.0){\rule[-0.200pt]{0.400pt}{4.818pt}}
\put(366,68){\makebox(0,0){2}}
\put(366.0,497.0){\rule[-0.200pt]{0.400pt}{4.818pt}}
\put(464.0,113.0){\rule[-0.200pt]{0.400pt}{4.818pt}}
\put(464,68){\makebox(0,0){3}}
\put(464.0,497.0){\rule[-0.200pt]{0.400pt}{4.818pt}}
\put(561.0,113.0){\rule[-0.200pt]{0.400pt}{4.818pt}}
\put(561,68){\makebox(0,0){4}}
\put(561.0,497.0){\rule[-0.200pt]{0.400pt}{4.818pt}}
\put(220.0,113.0){\rule[-0.200pt]{93.951pt}{0.400pt}}
\put(610.0,113.0){\rule[-0.200pt]{0.400pt}{97.324pt}}
\put(220.0,517.0){\rule[-0.200pt]{93.951pt}{0.400pt}}
\put(65,315){\makebox(0,0){$G_R$}}
\put(415,23){\makebox(0,0){$d$}}
\put(220.0,113.0){\rule[-0.200pt]{0.400pt}{97.324pt}}
\put(366,430){\raisebox{-.8pt}{\makebox(0,0){$\Diamond$}}}
\put(464,336){\raisebox{-.8pt}{\makebox(0,0){$\Diamond$}}}
\put(561,284){\raisebox{-.8pt}{\makebox(0,0){$\Diamond$}}}
\put(366.0,403.0){\rule[-0.200pt]{0.400pt}{13.009pt}}
\put(356.0,403.0){\rule[-0.200pt]{4.818pt}{0.400pt}}
\put(356.0,457.0){\rule[-0.200pt]{4.818pt}{0.400pt}}
\put(464.0,302.0){\rule[-0.200pt]{0.400pt}{16.622pt}}
\put(454.0,302.0){\rule[-0.200pt]{4.818pt}{0.400pt}}
\put(454.0,371.0){\rule[-0.200pt]{4.818pt}{0.400pt}}
\put(561.0,235.0){\rule[-0.200pt]{0.400pt}{23.367pt}}
\put(551.0,235.0){\rule[-0.200pt]{4.818pt}{0.400pt}}
\put(551.0,332.0){\rule[-0.200pt]{4.818pt}{0.400pt}}
\put(340.67,513){\rule{0.400pt}{0.964pt}}
\multiput(340.17,515.00)(1.000,-2.000){2}{\rule{0.400pt}{0.482pt}}
\multiput(342.60,505.53)(0.468,-2.382){5}{\rule{0.113pt}{1.800pt}}
\multiput(341.17,509.26)(4.000,-13.264){2}{\rule{0.400pt}{0.900pt}}
\multiput(346.60,488.94)(0.468,-2.236){5}{\rule{0.113pt}{1.700pt}}
\multiput(345.17,492.47)(4.000,-12.472){2}{\rule{0.400pt}{0.850pt}}
\multiput(350.60,473.77)(0.468,-1.943){5}{\rule{0.113pt}{1.500pt}}
\multiput(349.17,476.89)(4.000,-10.887){2}{\rule{0.400pt}{0.750pt}}
\multiput(354.60,460.19)(0.468,-1.797){5}{\rule{0.113pt}{1.400pt}}
\multiput(353.17,463.09)(4.000,-10.094){2}{\rule{0.400pt}{0.700pt}}
\multiput(358.60,447.60)(0.468,-1.651){5}{\rule{0.113pt}{1.300pt}}
\multiput(357.17,450.30)(4.000,-9.302){2}{\rule{0.400pt}{0.650pt}}
\multiput(362.60,436.43)(0.468,-1.358){5}{\rule{0.113pt}{1.100pt}}
\multiput(361.17,438.72)(4.000,-7.717){2}{\rule{0.400pt}{0.550pt}}
\multiput(366.60,426.43)(0.468,-1.358){5}{\rule{0.113pt}{1.100pt}}
\multiput(365.17,428.72)(4.000,-7.717){2}{\rule{0.400pt}{0.550pt}}
\multiput(370.60,416.85)(0.468,-1.212){5}{\rule{0.113pt}{1.000pt}}
\multiput(369.17,418.92)(4.000,-6.924){2}{\rule{0.400pt}{0.500pt}}
\multiput(374.60,408.68)(0.468,-0.920){5}{\rule{0.113pt}{0.800pt}}
\multiput(373.17,410.34)(4.000,-5.340){2}{\rule{0.400pt}{0.400pt}}
\multiput(378.60,401.26)(0.468,-1.066){5}{\rule{0.113pt}{0.900pt}}
\multiput(377.17,403.13)(4.000,-6.132){2}{\rule{0.400pt}{0.450pt}}
\multiput(382.61,393.26)(0.447,-1.132){3}{\rule{0.108pt}{0.900pt}}
\multiput(381.17,395.13)(3.000,-4.132){2}{\rule{0.400pt}{0.450pt}}
\multiput(385.60,388.09)(0.468,-0.774){5}{\rule{0.113pt}{0.700pt}}
\multiput(384.17,389.55)(4.000,-4.547){2}{\rule{0.400pt}{0.350pt}}
\multiput(389.60,382.09)(0.468,-0.774){5}{\rule{0.113pt}{0.700pt}}
\multiput(388.17,383.55)(4.000,-4.547){2}{\rule{0.400pt}{0.350pt}}
\multiput(393.60,376.51)(0.468,-0.627){5}{\rule{0.113pt}{0.600pt}}
\multiput(392.17,377.75)(4.000,-3.755){2}{\rule{0.400pt}{0.300pt}}
\multiput(397.00,372.94)(0.481,-0.468){5}{\rule{0.500pt}{0.113pt}}
\multiput(397.00,373.17)(2.962,-4.000){2}{\rule{0.250pt}{0.400pt}}
\multiput(401.60,367.51)(0.468,-0.627){5}{\rule{0.113pt}{0.600pt}}
\multiput(400.17,368.75)(4.000,-3.755){2}{\rule{0.400pt}{0.300pt}}
\multiput(405.00,363.95)(0.685,-0.447){3}{\rule{0.633pt}{0.108pt}}
\multiput(405.00,364.17)(2.685,-3.000){2}{\rule{0.317pt}{0.400pt}}
\multiput(409.00,360.94)(0.481,-0.468){5}{\rule{0.500pt}{0.113pt}}
\multiput(409.00,361.17)(2.962,-4.000){2}{\rule{0.250pt}{0.400pt}}
\multiput(413.00,356.95)(0.685,-0.447){3}{\rule{0.633pt}{0.108pt}}
\multiput(413.00,357.17)(2.685,-3.000){2}{\rule{0.317pt}{0.400pt}}
\multiput(417.00,353.95)(0.685,-0.447){3}{\rule{0.633pt}{0.108pt}}
\multiput(417.00,354.17)(2.685,-3.000){2}{\rule{0.317pt}{0.400pt}}
\multiput(421.00,350.95)(0.685,-0.447){3}{\rule{0.633pt}{0.108pt}}
\multiput(421.00,351.17)(2.685,-3.000){2}{\rule{0.317pt}{0.400pt}}
\multiput(425.00,347.95)(0.685,-0.447){3}{\rule{0.633pt}{0.108pt}}
\multiput(425.00,348.17)(2.685,-3.000){2}{\rule{0.317pt}{0.400pt}}
\put(429,344.17){\rule{0.900pt}{0.400pt}}
\multiput(429.00,345.17)(2.132,-2.000){2}{\rule{0.450pt}{0.400pt}}
\put(433,342.17){\rule{0.900pt}{0.400pt}}
\multiput(433.00,343.17)(2.132,-2.000){2}{\rule{0.450pt}{0.400pt}}
\put(437,340.17){\rule{0.900pt}{0.400pt}}
\multiput(437.00,341.17)(2.132,-2.000){2}{\rule{0.450pt}{0.400pt}}
\put(441,338.17){\rule{0.900pt}{0.400pt}}
\multiput(441.00,339.17)(2.132,-2.000){2}{\rule{0.450pt}{0.400pt}}
\put(445,336.67){\rule{0.723pt}{0.400pt}}
\multiput(445.00,337.17)(1.500,-1.000){2}{\rule{0.361pt}{0.400pt}}
\put(448,335.17){\rule{0.900pt}{0.400pt}}
\multiput(448.00,336.17)(2.132,-2.000){2}{\rule{0.450pt}{0.400pt}}
\put(452,333.67){\rule{0.964pt}{0.400pt}}
\multiput(452.00,334.17)(2.000,-1.000){2}{\rule{0.482pt}{0.400pt}}
\put(456,332.17){\rule{0.900pt}{0.400pt}}
\multiput(456.00,333.17)(2.132,-2.000){2}{\rule{0.450pt}{0.400pt}}
\put(460,330.67){\rule{0.964pt}{0.400pt}}
\multiput(460.00,331.17)(2.000,-1.000){2}{\rule{0.482pt}{0.400pt}}
\put(464,329.67){\rule{0.964pt}{0.400pt}}
\multiput(464.00,330.17)(2.000,-1.000){2}{\rule{0.482pt}{0.400pt}}
\put(468,328.67){\rule{0.964pt}{0.400pt}}
\multiput(468.00,329.17)(2.000,-1.000){2}{\rule{0.482pt}{0.400pt}}
\put(472,327.67){\rule{0.964pt}{0.400pt}}
\multiput(472.00,328.17)(2.000,-1.000){2}{\rule{0.482pt}{0.400pt}}
\put(476,326.67){\rule{0.964pt}{0.400pt}}
\multiput(476.00,327.17)(2.000,-1.000){2}{\rule{0.482pt}{0.400pt}}
\put(480,325.67){\rule{0.964pt}{0.400pt}}
\multiput(480.00,326.17)(2.000,-1.000){2}{\rule{0.482pt}{0.400pt}}
\put(484,324.67){\rule{0.964pt}{0.400pt}}
\multiput(484.00,325.17)(2.000,-1.000){2}{\rule{0.482pt}{0.400pt}}
\put(492,323.67){\rule{0.964pt}{0.400pt}}
\multiput(492.00,324.17)(2.000,-1.000){2}{\rule{0.482pt}{0.400pt}}
\put(496,322.67){\rule{0.964pt}{0.400pt}}
\multiput(496.00,323.17)(2.000,-1.000){2}{\rule{0.482pt}{0.400pt}}
\put(488.0,325.0){\rule[-0.200pt]{0.964pt}{0.400pt}}
\put(504,321.67){\rule{0.964pt}{0.400pt}}
\multiput(504.00,322.17)(2.000,-1.000){2}{\rule{0.482pt}{0.400pt}}
\put(500.0,323.0){\rule[-0.200pt]{0.964pt}{0.400pt}}
\put(512,320.67){\rule{0.723pt}{0.400pt}}
\multiput(512.00,321.17)(1.500,-1.000){2}{\rule{0.361pt}{0.400pt}}
\put(508.0,322.0){\rule[-0.200pt]{0.964pt}{0.400pt}}
\put(523,319.67){\rule{0.964pt}{0.400pt}}
\multiput(523.00,320.17)(2.000,-1.000){2}{\rule{0.482pt}{0.400pt}}
\put(515.0,321.0){\rule[-0.200pt]{1.927pt}{0.400pt}}
\put(535,318.67){\rule{0.964pt}{0.400pt}}
\multiput(535.00,319.17)(2.000,-1.000){2}{\rule{0.482pt}{0.400pt}}
\put(527.0,320.0){\rule[-0.200pt]{1.927pt}{0.400pt}}
\put(547,317.67){\rule{0.964pt}{0.400pt}}
\multiput(547.00,318.17)(2.000,-1.000){2}{\rule{0.482pt}{0.400pt}}
\put(539.0,319.0){\rule[-0.200pt]{1.927pt}{0.400pt}}
\put(567,316.67){\rule{0.964pt}{0.400pt}}
\multiput(567.00,317.17)(2.000,-1.000){2}{\rule{0.482pt}{0.400pt}}
\put(551.0,318.0){\rule[-0.200pt]{3.854pt}{0.400pt}}
\put(598,315.67){\rule{0.964pt}{0.400pt}}
\multiput(598.00,316.17)(2.000,-1.000){2}{\rule{0.482pt}{0.400pt}}
\put(571.0,317.0){\rule[-0.200pt]{6.504pt}{0.400pt}}
\put(602.0,316.0){\rule[-0.200pt]{1.927pt}{0.400pt}}
\put(366,442){\raisebox{-.8pt}{\makebox(0,0){$\Box$}}}
\put(464,343){\raisebox{-.8pt}{\makebox(0,0){$\Box$}}}
\put(561,274){\raisebox{-.8pt}{\makebox(0,0){$\Box$}}}
\put(366.0,428.0){\rule[-0.200pt]{0.400pt}{6.986pt}}
\put(356.0,428.0){\rule[-0.200pt]{4.818pt}{0.400pt}}
\put(356.0,457.0){\rule[-0.200pt]{4.818pt}{0.400pt}}
\put(464.0,324.0){\rule[-0.200pt]{0.400pt}{9.154pt}}
\put(454.0,324.0){\rule[-0.200pt]{4.818pt}{0.400pt}}
\put(454.0,362.0){\rule[-0.200pt]{4.818pt}{0.400pt}}
\put(561.0,240.0){\rule[-0.200pt]{0.400pt}{16.381pt}}
\put(551.0,240.0){\rule[-0.200pt]{4.818pt}{0.400pt}}
\put(551.0,308.0){\rule[-0.200pt]{4.818pt}{0.400pt}}
\put(344.17,507){\rule{0.400pt}{2.100pt}}
\multiput(343.17,512.64)(2.000,-5.641){2}{\rule{0.400pt}{1.050pt}}
\multiput(346.60,500.36)(0.468,-2.090){5}{\rule{0.113pt}{1.600pt}}
\multiput(345.17,503.68)(4.000,-11.679){2}{\rule{0.400pt}{0.800pt}}
\multiput(350.60,485.77)(0.468,-1.943){5}{\rule{0.113pt}{1.500pt}}
\multiput(349.17,488.89)(4.000,-10.887){2}{\rule{0.400pt}{0.750pt}}
\multiput(354.60,472.60)(0.468,-1.651){5}{\rule{0.113pt}{1.300pt}}
\multiput(353.17,475.30)(4.000,-9.302){2}{\rule{0.400pt}{0.650pt}}
\multiput(358.60,460.60)(0.468,-1.651){5}{\rule{0.113pt}{1.300pt}}
\multiput(357.17,463.30)(4.000,-9.302){2}{\rule{0.400pt}{0.650pt}}
\multiput(362.60,449.43)(0.468,-1.358){5}{\rule{0.113pt}{1.100pt}}
\multiput(361.17,451.72)(4.000,-7.717){2}{\rule{0.400pt}{0.550pt}}
\multiput(366.60,439.43)(0.468,-1.358){5}{\rule{0.113pt}{1.100pt}}
\multiput(365.17,441.72)(4.000,-7.717){2}{\rule{0.400pt}{0.550pt}}
\multiput(370.60,430.26)(0.468,-1.066){5}{\rule{0.113pt}{0.900pt}}
\multiput(369.17,432.13)(4.000,-6.132){2}{\rule{0.400pt}{0.450pt}}
\multiput(374.60,422.26)(0.468,-1.066){5}{\rule{0.113pt}{0.900pt}}
\multiput(373.17,424.13)(4.000,-6.132){2}{\rule{0.400pt}{0.450pt}}
\multiput(378.60,414.26)(0.468,-1.066){5}{\rule{0.113pt}{0.900pt}}
\multiput(377.17,416.13)(4.000,-6.132){2}{\rule{0.400pt}{0.450pt}}
\multiput(382.61,405.71)(0.447,-1.355){3}{\rule{0.108pt}{1.033pt}}
\multiput(381.17,407.86)(3.000,-4.855){2}{\rule{0.400pt}{0.517pt}}
\multiput(385.60,400.09)(0.468,-0.774){5}{\rule{0.113pt}{0.700pt}}
\multiput(384.17,401.55)(4.000,-4.547){2}{\rule{0.400pt}{0.350pt}}
\multiput(389.60,394.09)(0.468,-0.774){5}{\rule{0.113pt}{0.700pt}}
\multiput(388.17,395.55)(4.000,-4.547){2}{\rule{0.400pt}{0.350pt}}
\multiput(393.60,388.51)(0.468,-0.627){5}{\rule{0.113pt}{0.600pt}}
\multiput(392.17,389.75)(4.000,-3.755){2}{\rule{0.400pt}{0.300pt}}
\multiput(397.60,383.51)(0.468,-0.627){5}{\rule{0.113pt}{0.600pt}}
\multiput(396.17,384.75)(4.000,-3.755){2}{\rule{0.400pt}{0.300pt}}
\multiput(401.00,379.94)(0.481,-0.468){5}{\rule{0.500pt}{0.113pt}}
\multiput(401.00,380.17)(2.962,-4.000){2}{\rule{0.250pt}{0.400pt}}
\multiput(405.60,374.51)(0.468,-0.627){5}{\rule{0.113pt}{0.600pt}}
\multiput(404.17,375.75)(4.000,-3.755){2}{\rule{0.400pt}{0.300pt}}
\multiput(409.00,370.95)(0.685,-0.447){3}{\rule{0.633pt}{0.108pt}}
\multiput(409.00,371.17)(2.685,-3.000){2}{\rule{0.317pt}{0.400pt}}
\multiput(413.00,367.94)(0.481,-0.468){5}{\rule{0.500pt}{0.113pt}}
\multiput(413.00,368.17)(2.962,-4.000){2}{\rule{0.250pt}{0.400pt}}
\multiput(417.00,363.95)(0.685,-0.447){3}{\rule{0.633pt}{0.108pt}}
\multiput(417.00,364.17)(2.685,-3.000){2}{\rule{0.317pt}{0.400pt}}
\multiput(421.00,360.95)(0.685,-0.447){3}{\rule{0.633pt}{0.108pt}}
\multiput(421.00,361.17)(2.685,-3.000){2}{\rule{0.317pt}{0.400pt}}
\multiput(425.00,357.95)(0.685,-0.447){3}{\rule{0.633pt}{0.108pt}}
\multiput(425.00,358.17)(2.685,-3.000){2}{\rule{0.317pt}{0.400pt}}
\multiput(429.00,354.95)(0.685,-0.447){3}{\rule{0.633pt}{0.108pt}}
\multiput(429.00,355.17)(2.685,-3.000){2}{\rule{0.317pt}{0.400pt}}
\put(433,351.17){\rule{0.900pt}{0.400pt}}
\multiput(433.00,352.17)(2.132,-2.000){2}{\rule{0.450pt}{0.400pt}}
\multiput(437.00,349.95)(0.685,-0.447){3}{\rule{0.633pt}{0.108pt}}
\multiput(437.00,350.17)(2.685,-3.000){2}{\rule{0.317pt}{0.400pt}}
\put(441,346.17){\rule{0.900pt}{0.400pt}}
\multiput(441.00,347.17)(2.132,-2.000){2}{\rule{0.450pt}{0.400pt}}
\put(445,344.17){\rule{0.700pt}{0.400pt}}
\multiput(445.00,345.17)(1.547,-2.000){2}{\rule{0.350pt}{0.400pt}}
\put(448,342.17){\rule{0.900pt}{0.400pt}}
\multiput(448.00,343.17)(2.132,-2.000){2}{\rule{0.450pt}{0.400pt}}
\put(452,340.67){\rule{0.964pt}{0.400pt}}
\multiput(452.00,341.17)(2.000,-1.000){2}{\rule{0.482pt}{0.400pt}}
\put(456,339.17){\rule{0.900pt}{0.400pt}}
\multiput(456.00,340.17)(2.132,-2.000){2}{\rule{0.450pt}{0.400pt}}
\put(460,337.67){\rule{0.964pt}{0.400pt}}
\multiput(460.00,338.17)(2.000,-1.000){2}{\rule{0.482pt}{0.400pt}}
\put(464,336.17){\rule{0.900pt}{0.400pt}}
\multiput(464.00,337.17)(2.132,-2.000){2}{\rule{0.450pt}{0.400pt}}
\put(468,334.67){\rule{0.964pt}{0.400pt}}
\multiput(468.00,335.17)(2.000,-1.000){2}{\rule{0.482pt}{0.400pt}}
\put(472,333.67){\rule{0.964pt}{0.400pt}}
\multiput(472.00,334.17)(2.000,-1.000){2}{\rule{0.482pt}{0.400pt}}
\put(476,332.67){\rule{0.964pt}{0.400pt}}
\multiput(476.00,333.17)(2.000,-1.000){2}{\rule{0.482pt}{0.400pt}}
\put(480,331.67){\rule{0.964pt}{0.400pt}}
\multiput(480.00,332.17)(2.000,-1.000){2}{\rule{0.482pt}{0.400pt}}
\put(484,330.67){\rule{0.964pt}{0.400pt}}
\multiput(484.00,331.17)(2.000,-1.000){2}{\rule{0.482pt}{0.400pt}}
\put(488,329.67){\rule{0.964pt}{0.400pt}}
\multiput(488.00,330.17)(2.000,-1.000){2}{\rule{0.482pt}{0.400pt}}
\put(492,328.67){\rule{0.964pt}{0.400pt}}
\multiput(492.00,329.17)(2.000,-1.000){2}{\rule{0.482pt}{0.400pt}}
\put(496,327.67){\rule{0.964pt}{0.400pt}}
\multiput(496.00,328.17)(2.000,-1.000){2}{\rule{0.482pt}{0.400pt}}
\put(500,326.67){\rule{0.964pt}{0.400pt}}
\multiput(500.00,327.17)(2.000,-1.000){2}{\rule{0.482pt}{0.400pt}}
\put(504,325.67){\rule{0.964pt}{0.400pt}}
\multiput(504.00,326.17)(2.000,-1.000){2}{\rule{0.482pt}{0.400pt}}
\put(512,324.67){\rule{0.723pt}{0.400pt}}
\multiput(512.00,325.17)(1.500,-1.000){2}{\rule{0.361pt}{0.400pt}}
\put(508.0,326.0){\rule[-0.200pt]{0.964pt}{0.400pt}}
\put(519,323.67){\rule{0.964pt}{0.400pt}}
\multiput(519.00,324.17)(2.000,-1.000){2}{\rule{0.482pt}{0.400pt}}
\put(523,322.67){\rule{0.964pt}{0.400pt}}
\multiput(523.00,323.17)(2.000,-1.000){2}{\rule{0.482pt}{0.400pt}}
\put(515.0,325.0){\rule[-0.200pt]{0.964pt}{0.400pt}}
\put(535,321.67){\rule{0.964pt}{0.400pt}}
\multiput(535.00,322.17)(2.000,-1.000){2}{\rule{0.482pt}{0.400pt}}
\put(527.0,323.0){\rule[-0.200pt]{1.927pt}{0.400pt}}
\put(543,320.67){\rule{0.964pt}{0.400pt}}
\multiput(543.00,321.17)(2.000,-1.000){2}{\rule{0.482pt}{0.400pt}}
\put(539.0,322.0){\rule[-0.200pt]{0.964pt}{0.400pt}}
\put(555,319.67){\rule{0.964pt}{0.400pt}}
\multiput(555.00,320.17)(2.000,-1.000){2}{\rule{0.482pt}{0.400pt}}
\put(547.0,321.0){\rule[-0.200pt]{1.927pt}{0.400pt}}
\put(571,318.67){\rule{0.964pt}{0.400pt}}
\multiput(571.00,319.17)(2.000,-1.000){2}{\rule{0.482pt}{0.400pt}}
\put(559.0,320.0){\rule[-0.200pt]{2.891pt}{0.400pt}}
\put(586,317.67){\rule{0.964pt}{0.400pt}}
\multiput(586.00,318.17)(2.000,-1.000){2}{\rule{0.482pt}{0.400pt}}
\put(575.0,319.0){\rule[-0.200pt]{2.650pt}{0.400pt}}
\put(590.0,318.0){\rule[-0.200pt]{4.818pt}{0.400pt}}
\put(366,416){\makebox(0,0){$\triangle$}}
\put(464,370){\makebox(0,0){$\triangle$}}
\put(561,264){\makebox(0,0){$\triangle$}}
\put(366.0,397.0){\rule[-0.200pt]{0.400pt}{9.395pt}}
\put(356.0,397.0){\rule[-0.200pt]{4.818pt}{0.400pt}}
\put(356.0,436.0){\rule[-0.200pt]{4.818pt}{0.400pt}}
\put(464.0,347.0){\rule[-0.200pt]{0.400pt}{11.322pt}}
\put(454.0,347.0){\rule[-0.200pt]{4.818pt}{0.400pt}}
\put(454.0,394.0){\rule[-0.200pt]{4.818pt}{0.400pt}}
\put(561.0,234.0){\rule[-0.200pt]{0.400pt}{14.213pt}}
\put(551.0,234.0){\rule[-0.200pt]{4.818pt}{0.400pt}}
\put(551.0,293.0){\rule[-0.200pt]{4.818pt}{0.400pt}}
\multiput(327.61,509.39)(0.447,-2.695){3}{\rule{0.108pt}{1.833pt}}
\multiput(326.17,513.19)(3.000,-9.195){2}{\rule{0.400pt}{0.917pt}}
\multiput(330.60,498.19)(0.468,-1.797){5}{\rule{0.113pt}{1.400pt}}
\multiput(329.17,501.09)(4.000,-10.094){2}{\rule{0.400pt}{0.700pt}}
\multiput(334.60,485.60)(0.468,-1.651){5}{\rule{0.113pt}{1.300pt}}
\multiput(333.17,488.30)(4.000,-9.302){2}{\rule{0.400pt}{0.650pt}}
\multiput(338.60,474.02)(0.468,-1.505){5}{\rule{0.113pt}{1.200pt}}
\multiput(337.17,476.51)(4.000,-8.509){2}{\rule{0.400pt}{0.600pt}}
\multiput(342.60,463.43)(0.468,-1.358){5}{\rule{0.113pt}{1.100pt}}
\multiput(341.17,465.72)(4.000,-7.717){2}{\rule{0.400pt}{0.550pt}}
\multiput(346.60,453.85)(0.468,-1.212){5}{\rule{0.113pt}{1.000pt}}
\multiput(345.17,455.92)(4.000,-6.924){2}{\rule{0.400pt}{0.500pt}}
\multiput(350.60,445.26)(0.468,-1.066){5}{\rule{0.113pt}{0.900pt}}
\multiput(349.17,447.13)(4.000,-6.132){2}{\rule{0.400pt}{0.450pt}}
\multiput(354.60,437.26)(0.468,-1.066){5}{\rule{0.113pt}{0.900pt}}
\multiput(353.17,439.13)(4.000,-6.132){2}{\rule{0.400pt}{0.450pt}}
\multiput(358.60,429.68)(0.468,-0.920){5}{\rule{0.113pt}{0.800pt}}
\multiput(357.17,431.34)(4.000,-5.340){2}{\rule{0.400pt}{0.400pt}}
\multiput(362.60,422.68)(0.468,-0.920){5}{\rule{0.113pt}{0.800pt}}
\multiput(361.17,424.34)(4.000,-5.340){2}{\rule{0.400pt}{0.400pt}}
\multiput(366.60,416.09)(0.468,-0.774){5}{\rule{0.113pt}{0.700pt}}
\multiput(365.17,417.55)(4.000,-4.547){2}{\rule{0.400pt}{0.350pt}}
\multiput(370.60,410.09)(0.468,-0.774){5}{\rule{0.113pt}{0.700pt}}
\multiput(369.17,411.55)(4.000,-4.547){2}{\rule{0.400pt}{0.350pt}}
\multiput(374.60,404.51)(0.468,-0.627){5}{\rule{0.113pt}{0.600pt}}
\multiput(373.17,405.75)(4.000,-3.755){2}{\rule{0.400pt}{0.300pt}}
\multiput(378.60,399.51)(0.468,-0.627){5}{\rule{0.113pt}{0.600pt}}
\multiput(377.17,400.75)(4.000,-3.755){2}{\rule{0.400pt}{0.300pt}}
\multiput(382.61,393.82)(0.447,-0.909){3}{\rule{0.108pt}{0.767pt}}
\multiput(381.17,395.41)(3.000,-3.409){2}{\rule{0.400pt}{0.383pt}}
\multiput(385.00,390.94)(0.481,-0.468){5}{\rule{0.500pt}{0.113pt}}
\multiput(385.00,391.17)(2.962,-4.000){2}{\rule{0.250pt}{0.400pt}}
\multiput(389.00,386.94)(0.481,-0.468){5}{\rule{0.500pt}{0.113pt}}
\multiput(389.00,387.17)(2.962,-4.000){2}{\rule{0.250pt}{0.400pt}}
\multiput(393.00,382.94)(0.481,-0.468){5}{\rule{0.500pt}{0.113pt}}
\multiput(393.00,383.17)(2.962,-4.000){2}{\rule{0.250pt}{0.400pt}}
\multiput(397.00,378.94)(0.481,-0.468){5}{\rule{0.500pt}{0.113pt}}
\multiput(397.00,379.17)(2.962,-4.000){2}{\rule{0.250pt}{0.400pt}}
\multiput(401.00,374.95)(0.685,-0.447){3}{\rule{0.633pt}{0.108pt}}
\multiput(401.00,375.17)(2.685,-3.000){2}{\rule{0.317pt}{0.400pt}}
\multiput(405.00,371.95)(0.685,-0.447){3}{\rule{0.633pt}{0.108pt}}
\multiput(405.00,372.17)(2.685,-3.000){2}{\rule{0.317pt}{0.400pt}}
\multiput(409.00,368.95)(0.685,-0.447){3}{\rule{0.633pt}{0.108pt}}
\multiput(409.00,369.17)(2.685,-3.000){2}{\rule{0.317pt}{0.400pt}}
\multiput(413.00,365.95)(0.685,-0.447){3}{\rule{0.633pt}{0.108pt}}
\multiput(413.00,366.17)(2.685,-3.000){2}{\rule{0.317pt}{0.400pt}}
\put(417,362.17){\rule{0.900pt}{0.400pt}}
\multiput(417.00,363.17)(2.132,-2.000){2}{\rule{0.450pt}{0.400pt}}
\multiput(421.00,360.95)(0.685,-0.447){3}{\rule{0.633pt}{0.108pt}}
\multiput(421.00,361.17)(2.685,-3.000){2}{\rule{0.317pt}{0.400pt}}
\put(425,357.17){\rule{0.900pt}{0.400pt}}
\multiput(425.00,358.17)(2.132,-2.000){2}{\rule{0.450pt}{0.400pt}}
\put(429,355.17){\rule{0.900pt}{0.400pt}}
\multiput(429.00,356.17)(2.132,-2.000){2}{\rule{0.450pt}{0.400pt}}
\put(433,353.17){\rule{0.900pt}{0.400pt}}
\multiput(433.00,354.17)(2.132,-2.000){2}{\rule{0.450pt}{0.400pt}}
\put(437,351.17){\rule{0.900pt}{0.400pt}}
\multiput(437.00,352.17)(2.132,-2.000){2}{\rule{0.450pt}{0.400pt}}
\put(441,349.17){\rule{0.900pt}{0.400pt}}
\multiput(441.00,350.17)(2.132,-2.000){2}{\rule{0.450pt}{0.400pt}}
\put(445,347.67){\rule{0.723pt}{0.400pt}}
\multiput(445.00,348.17)(1.500,-1.000){2}{\rule{0.361pt}{0.400pt}}
\put(448,346.17){\rule{0.900pt}{0.400pt}}
\multiput(448.00,347.17)(2.132,-2.000){2}{\rule{0.450pt}{0.400pt}}
\put(452,344.67){\rule{0.964pt}{0.400pt}}
\multiput(452.00,345.17)(2.000,-1.000){2}{\rule{0.482pt}{0.400pt}}
\put(456,343.17){\rule{0.900pt}{0.400pt}}
\multiput(456.00,344.17)(2.132,-2.000){2}{\rule{0.450pt}{0.400pt}}
\put(460,341.67){\rule{0.964pt}{0.400pt}}
\multiput(460.00,342.17)(2.000,-1.000){2}{\rule{0.482pt}{0.400pt}}
\put(464,340.67){\rule{0.964pt}{0.400pt}}
\multiput(464.00,341.17)(2.000,-1.000){2}{\rule{0.482pt}{0.400pt}}
\put(468,339.17){\rule{0.900pt}{0.400pt}}
\multiput(468.00,340.17)(2.132,-2.000){2}{\rule{0.450pt}{0.400pt}}
\put(472,337.67){\rule{0.964pt}{0.400pt}}
\multiput(472.00,338.17)(2.000,-1.000){2}{\rule{0.482pt}{0.400pt}}
\put(476,336.67){\rule{0.964pt}{0.400pt}}
\multiput(476.00,337.17)(2.000,-1.000){2}{\rule{0.482pt}{0.400pt}}
\put(480,335.67){\rule{0.964pt}{0.400pt}}
\multiput(480.00,336.17)(2.000,-1.000){2}{\rule{0.482pt}{0.400pt}}
\put(484,334.67){\rule{0.964pt}{0.400pt}}
\multiput(484.00,335.17)(2.000,-1.000){2}{\rule{0.482pt}{0.400pt}}
\put(488,333.67){\rule{0.964pt}{0.400pt}}
\multiput(488.00,334.17)(2.000,-1.000){2}{\rule{0.482pt}{0.400pt}}
\put(492,332.67){\rule{0.964pt}{0.400pt}}
\multiput(492.00,333.17)(2.000,-1.000){2}{\rule{0.482pt}{0.400pt}}
\put(500,331.67){\rule{0.964pt}{0.400pt}}
\multiput(500.00,332.17)(2.000,-1.000){2}{\rule{0.482pt}{0.400pt}}
\put(504,330.67){\rule{0.964pt}{0.400pt}}
\multiput(504.00,331.17)(2.000,-1.000){2}{\rule{0.482pt}{0.400pt}}
\put(508,329.67){\rule{0.964pt}{0.400pt}}
\multiput(508.00,330.17)(2.000,-1.000){2}{\rule{0.482pt}{0.400pt}}
\put(496.0,333.0){\rule[-0.200pt]{0.964pt}{0.400pt}}
\put(515,328.67){\rule{0.964pt}{0.400pt}}
\multiput(515.00,329.17)(2.000,-1.000){2}{\rule{0.482pt}{0.400pt}}
\put(512.0,330.0){\rule[-0.200pt]{0.723pt}{0.400pt}}
\put(523,327.67){\rule{0.964pt}{0.400pt}}
\multiput(523.00,328.17)(2.000,-1.000){2}{\rule{0.482pt}{0.400pt}}
\put(527,326.67){\rule{0.964pt}{0.400pt}}
\multiput(527.00,327.17)(2.000,-1.000){2}{\rule{0.482pt}{0.400pt}}
\put(519.0,329.0){\rule[-0.200pt]{0.964pt}{0.400pt}}
\put(535,325.67){\rule{0.964pt}{0.400pt}}
\multiput(535.00,326.17)(2.000,-1.000){2}{\rule{0.482pt}{0.400pt}}
\put(531.0,327.0){\rule[-0.200pt]{0.964pt}{0.400pt}}
\put(543,324.67){\rule{0.964pt}{0.400pt}}
\multiput(543.00,325.17)(2.000,-1.000){2}{\rule{0.482pt}{0.400pt}}
\put(539.0,326.0){\rule[-0.200pt]{0.964pt}{0.400pt}}
\put(555,323.67){\rule{0.964pt}{0.400pt}}
\multiput(555.00,324.17)(2.000,-1.000){2}{\rule{0.482pt}{0.400pt}}
\put(547.0,325.0){\rule[-0.200pt]{1.927pt}{0.400pt}}
\put(563,322.67){\rule{0.964pt}{0.400pt}}
\multiput(563.00,323.17)(2.000,-1.000){2}{\rule{0.482pt}{0.400pt}}
\put(559.0,324.0){\rule[-0.200pt]{0.964pt}{0.400pt}}
\put(575,321.67){\rule{0.723pt}{0.400pt}}
\multiput(575.00,322.17)(1.500,-1.000){2}{\rule{0.361pt}{0.400pt}}
\put(567.0,323.0){\rule[-0.200pt]{1.927pt}{0.400pt}}
\put(590,320.67){\rule{0.964pt}{0.400pt}}
\multiput(590.00,321.17)(2.000,-1.000){2}{\rule{0.482pt}{0.400pt}}
\put(578.0,322.0){\rule[-0.200pt]{2.891pt}{0.400pt}}
\put(606,319.67){\rule{0.964pt}{0.400pt}}
\multiput(606.00,320.17)(2.000,-1.000){2}{\rule{0.482pt}{0.400pt}}
\put(594.0,321.0){\rule[-0.200pt]{2.891pt}{0.400pt}}
\end{picture}

%% file: fig31.tex
\setlength{\unitlength}{0.240900pt}
\ifx\plotpoint\undefined\newsavebox{\plotpoint}\fi
\sbox{\plotpoint}{\rule[-0.200pt]{0.400pt}{0.400pt}}%
\begin{picture}(600,540)(0,0)
\font\gnuplot=cmr10 at 10pt
\gnuplot
\sbox{\plotpoint}{\rule[-0.200pt]{0.400pt}{0.400pt}}%
\put(220.0,171.0){\rule[-0.200pt]{93.951pt}{0.400pt}}
\put(220.0,113.0){\rule[-0.200pt]{4.818pt}{0.400pt}}
\put(198,113){\makebox(0,0)[r]{-0.005}}
\put(590.0,113.0){\rule[-0.200pt]{4.818pt}{0.400pt}}
\put(220.0,171.0){\rule[-0.200pt]{4.818pt}{0.400pt}}
\put(198,171){\makebox(0,0)[r]{0}}
\put(590.0,171.0){\rule[-0.200pt]{4.818pt}{0.400pt}}
\put(220.0,228.0){\rule[-0.200pt]{4.818pt}{0.400pt}}
\put(198,228){\makebox(0,0)[r]{0.005}}
\put(590.0,228.0){\rule[-0.200pt]{4.818pt}{0.400pt}}
\put(220.0,286.0){\rule[-0.200pt]{4.818pt}{0.400pt}}
\put(198,286){\makebox(0,0)[r]{0.01}}
\put(590.0,286.0){\rule[-0.200pt]{4.818pt}{0.400pt}}
\put(220.0,344.0){\rule[-0.200pt]{4.818pt}{0.400pt}}
\put(198,344){\makebox(0,0)[r]{0.015}}
\put(590.0,344.0){\rule[-0.200pt]{4.818pt}{0.400pt}}
\put(220.0,402.0){\rule[-0.200pt]{4.818pt}{0.400pt}}
\put(198,402){\makebox(0,0)[r]{0.02}}
\put(590.0,402.0){\rule[-0.200pt]{4.818pt}{0.400pt}}
\put(220.0,459.0){\rule[-0.200pt]{4.818pt}{0.400pt}}
\put(198,459){\makebox(0,0)[r]{0.025}}
\put(590.0,459.0){\rule[-0.200pt]{4.818pt}{0.400pt}}
\put(220.0,517.0){\rule[-0.200pt]{4.818pt}{0.400pt}}
\put(198,517){\makebox(0,0)[r]{0.03}}
\put(590.0,517.0){\rule[-0.200pt]{4.818pt}{0.400pt}}
\put(269.0,113.0){\rule[-0.200pt]{0.400pt}{4.818pt}}
\put(269,68){\makebox(0,0){1}}
\put(269.0,497.0){\rule[-0.200pt]{0.400pt}{4.818pt}}
\put(366.0,113.0){\rule[-0.200pt]{0.400pt}{4.818pt}}
\put(366,68){\makebox(0,0){2}}
\put(366.0,497.0){\rule[-0.200pt]{0.400pt}{4.818pt}}
\put(464.0,113.0){\rule[-0.200pt]{0.400pt}{4.818pt}}
\put(464,68){\makebox(0,0){3}}
\put(464.0,497.0){\rule[-0.200pt]{0.400pt}{4.818pt}}
\put(561.0,113.0){\rule[-0.200pt]{0.400pt}{4.818pt}}
\put(561,68){\makebox(0,0){4}}
\put(561.0,497.0){\rule[-0.200pt]{0.400pt}{4.818pt}}
\put(220.0,113.0){\rule[-0.200pt]{93.951pt}{0.400pt}}
\put(610.0,113.0){\rule[-0.200pt]{0.400pt}{97.324pt}}
\put(220.0,517.0){\rule[-0.200pt]{93.951pt}{0.400pt}}
\put(45,315){\makebox(0,0){$G_V$}}
\put(415,23){\makebox(0,0){$d$}}
\put(220.0,113.0){\rule[-0.200pt]{0.400pt}{97.324pt}}
\put(269,501){\raisebox{-.8pt}{\makebox(0,0){$\Diamond$}}}
\put(366,210){\raisebox{-.8pt}{\makebox(0,0){$\Diamond$}}}
\put(464,168){\raisebox{-.8pt}{\makebox(0,0){$\Diamond$}}}
\put(561,144){\raisebox{-.8pt}{\makebox(0,0){$\Diamond$}}}
\put(269,427){\raisebox{-.8pt}{\makebox(0,0){$\Box$}}}
\put(366,174){\raisebox{-.8pt}{\makebox(0,0){$\Box$}}}
\put(464,165){\raisebox{-.8pt}{\makebox(0,0){$\Box$}}}
\put(561,161){\raisebox{-.8pt}{\makebox(0,0){$\Box$}}}
\end{picture}

%% file: fig32.tex
\setlength{\unitlength}{0.240900pt}
\ifx\plotpoint\undefined\newsavebox{\plotpoint}\fi
\sbox{\plotpoint}{\rule[-0.200pt]{0.400pt}{0.400pt}}%
\begin{picture}(580,540)(0,0)
\font\gnuplot=cmr10 at 10pt
\gnuplot
\sbox{\plotpoint}{\rule[-0.200pt]{0.400pt}{0.400pt}}%
\put(220.0,450.0){\rule[-0.200pt]{93.951pt}{0.400pt}}
\put(220.0,113.0){\rule[-0.200pt]{4.818pt}{0.400pt}}
\put(198,113){\makebox(0,0)[r]{-1}}
\put(590.0,113.0){\rule[-0.200pt]{4.818pt}{0.400pt}}
\put(220.0,180.0){\rule[-0.200pt]{4.818pt}{0.400pt}}
\put(198,180){\makebox(0,0)[r]{-0.8}}
\put(590.0,180.0){\rule[-0.200pt]{4.818pt}{0.400pt}}
\put(220.0,248.0){\rule[-0.200pt]{4.818pt}{0.400pt}}
\put(198,248){\makebox(0,0)[r]{-0.6}}
\put(590.0,248.0){\rule[-0.200pt]{4.818pt}{0.400pt}}
\put(220.0,315.0){\rule[-0.200pt]{4.818pt}{0.400pt}}
\put(198,315){\makebox(0,0)[r]{-0.4}}
\put(590.0,315.0){\rule[-0.200pt]{4.818pt}{0.400pt}}
\put(220.0,382.0){\rule[-0.200pt]{4.818pt}{0.400pt}}
\put(198,382){\makebox(0,0)[r]{-0.2}}
\put(590.0,382.0){\rule[-0.200pt]{4.818pt}{0.400pt}}
\put(220.0,450.0){\rule[-0.200pt]{4.818pt}{0.400pt}}
\put(198,450){\makebox(0,0)[r]{0}}
\put(590.0,450.0){\rule[-0.200pt]{4.818pt}{0.400pt}}
\put(220.0,517.0){\rule[-0.200pt]{4.818pt}{0.400pt}}
\put(198,517){\makebox(0,0)[r]{0.2}}
\put(590.0,517.0){\rule[-0.200pt]{4.818pt}{0.400pt}}
\put(269.0,113.0){\rule[-0.200pt]{0.400pt}{4.818pt}}
\put(269,68){\makebox(0,0){1}}
\put(269.0,497.0){\rule[-0.200pt]{0.400pt}{4.818pt}}
\put(366.0,113.0){\rule[-0.200pt]{0.400pt}{4.818pt}}
\put(366,68){\makebox(0,0){2}}
\put(366.0,497.0){\rule[-0.200pt]{0.400pt}{4.818pt}}
\put(464.0,113.0){\rule[-0.200pt]{0.400pt}{4.818pt}}
\put(464,68){\makebox(0,0){3}}
\put(464.0,497.0){\rule[-0.200pt]{0.400pt}{4.818pt}}
\put(561.0,113.0){\rule[-0.200pt]{0.400pt}{4.818pt}}
\put(561,68){\makebox(0,0){4}}
\put(561.0,497.0){\rule[-0.200pt]{0.400pt}{4.818pt}}
\put(220.0,113.0){\rule[-0.200pt]{93.951pt}{0.400pt}}
\put(610.0,113.0){\rule[-0.200pt]{0.400pt}{97.324pt}}
\put(220.0,517.0){\rule[-0.200pt]{93.951pt}{0.400pt}}
\put(65,315){\makebox(0,0){$G_R$}}
\put(415,23){\makebox(0,0){$d$}}
\put(220.0,113.0){\rule[-0.200pt]{0.400pt}{97.324pt}}
\put(269,218){\raisebox{-.8pt}{\makebox(0,0){$\Diamond$}}}
\put(366,456){\raisebox{-.8pt}{\makebox(0,0){$\Diamond$}}}
\put(464,443){\raisebox{-.8pt}{\makebox(0,0){$\Diamond$}}}
\put(561,428){\raisebox{-.8pt}{\makebox(0,0){$\Diamond$}}}
\put(297.67,515){\rule{0.400pt}{0.482pt}}
\multiput(297.17,516.00)(1.000,-1.000){2}{\rule{0.400pt}{0.241pt}}
\multiput(299.60,510.85)(0.468,-1.212){5}{\rule{0.113pt}{1.000pt}}
\multiput(298.17,512.92)(4.000,-6.924){2}{\rule{0.400pt}{0.500pt}}
\multiput(303.60,502.68)(0.468,-0.920){5}{\rule{0.113pt}{0.800pt}}
\multiput(302.17,504.34)(4.000,-5.340){2}{\rule{0.400pt}{0.400pt}}
\multiput(307.60,495.68)(0.468,-0.920){5}{\rule{0.113pt}{0.800pt}}
\multiput(306.17,497.34)(4.000,-5.340){2}{\rule{0.400pt}{0.400pt}}
\multiput(311.60,489.09)(0.468,-0.774){5}{\rule{0.113pt}{0.700pt}}
\multiput(310.17,490.55)(4.000,-4.547){2}{\rule{0.400pt}{0.350pt}}
\multiput(315.61,482.82)(0.447,-0.909){3}{\rule{0.108pt}{0.767pt}}
\multiput(314.17,484.41)(3.000,-3.409){2}{\rule{0.400pt}{0.383pt}}
\multiput(318.00,479.94)(0.481,-0.468){5}{\rule{0.500pt}{0.113pt}}
\multiput(318.00,480.17)(2.962,-4.000){2}{\rule{0.250pt}{0.400pt}}
\multiput(322.00,475.95)(0.685,-0.447){3}{\rule{0.633pt}{0.108pt}}
\multiput(322.00,476.17)(2.685,-3.000){2}{\rule{0.317pt}{0.400pt}}
\multiput(326.00,472.95)(0.685,-0.447){3}{\rule{0.633pt}{0.108pt}}
\multiput(326.00,473.17)(2.685,-3.000){2}{\rule{0.317pt}{0.400pt}}
\multiput(330.00,469.95)(0.685,-0.447){3}{\rule{0.633pt}{0.108pt}}
\multiput(330.00,470.17)(2.685,-3.000){2}{\rule{0.317pt}{0.400pt}}
\put(334,466.17){\rule{0.900pt}{0.400pt}}
\multiput(334.00,467.17)(2.132,-2.000){2}{\rule{0.450pt}{0.400pt}}
\put(338,464.17){\rule{0.900pt}{0.400pt}}
\multiput(338.00,465.17)(2.132,-2.000){2}{\rule{0.450pt}{0.400pt}}
\put(342,462.17){\rule{0.900pt}{0.400pt}}
\multiput(342.00,463.17)(2.132,-2.000){2}{\rule{0.450pt}{0.400pt}}
\put(346,460.17){\rule{0.900pt}{0.400pt}}
\multiput(346.00,461.17)(2.132,-2.000){2}{\rule{0.450pt}{0.400pt}}
\put(350,458.67){\rule{0.964pt}{0.400pt}}
\multiput(350.00,459.17)(2.000,-1.000){2}{\rule{0.482pt}{0.400pt}}
\put(354,457.67){\rule{0.964pt}{0.400pt}}
\multiput(354.00,458.17)(2.000,-1.000){2}{\rule{0.482pt}{0.400pt}}
\put(358,456.67){\rule{0.964pt}{0.400pt}}
\multiput(358.00,457.17)(2.000,-1.000){2}{\rule{0.482pt}{0.400pt}}
\put(362,455.67){\rule{0.964pt}{0.400pt}}
\multiput(362.00,456.17)(2.000,-1.000){2}{\rule{0.482pt}{0.400pt}}
\put(366,454.67){\rule{0.964pt}{0.400pt}}
\multiput(366.00,455.17)(2.000,-1.000){2}{\rule{0.482pt}{0.400pt}}
\put(374,453.67){\rule{0.964pt}{0.400pt}}
\multiput(374.00,454.17)(2.000,-1.000){2}{\rule{0.482pt}{0.400pt}}
\put(370.0,455.0){\rule[-0.200pt]{0.964pt}{0.400pt}}
\put(382,452.67){\rule{0.723pt}{0.400pt}}
\multiput(382.00,453.17)(1.500,-1.000){2}{\rule{0.361pt}{0.400pt}}
\put(378.0,454.0){\rule[-0.200pt]{0.964pt}{0.400pt}}
\put(389,451.67){\rule{0.964pt}{0.400pt}}
\multiput(389.00,452.17)(2.000,-1.000){2}{\rule{0.482pt}{0.400pt}}
\put(385.0,453.0){\rule[-0.200pt]{0.964pt}{0.400pt}}
\put(405,450.67){\rule{0.964pt}{0.400pt}}
\multiput(405.00,451.17)(2.000,-1.000){2}{\rule{0.482pt}{0.400pt}}
\put(393.0,452.0){\rule[-0.200pt]{2.891pt}{0.400pt}}
\put(429,449.67){\rule{0.964pt}{0.400pt}}
\multiput(429.00,450.17)(2.000,-1.000){2}{\rule{0.482pt}{0.400pt}}
\put(409.0,451.0){\rule[-0.200pt]{4.818pt}{0.400pt}}
\put(433.0,450.0){\rule[-0.200pt]{42.639pt}{0.400pt}}
\put(269,313){\raisebox{-.8pt}{\makebox(0,0){$\Box$}}}
\put(366,448){\raisebox{-.8pt}{\makebox(0,0){$\Box$}}}
\put(464,445){\raisebox{-.8pt}{\makebox(0,0){$\Box$}}}
\put(561,459){\raisebox{-.8pt}{\makebox(0,0){$\Box$}}}
\end{picture}

%% file: fig33.tex
\setlength{\unitlength}{0.240900pt}
\ifx\plotpoint\undefined\newsavebox{\plotpoint}\fi
\sbox{\plotpoint}{\rule[-0.200pt]{0.400pt}{0.400pt}}%
\begin{picture}(580,540)(0,0)
\font\gnuplot=cmr10 at 10pt
\gnuplot
\sbox{\plotpoint}{\rule[-0.200pt]{0.400pt}{0.400pt}}%
\put(220.0,315.0){\rule[-0.200pt]{93.951pt}{0.400pt}}
\put(220.0,113.0){\rule[-0.200pt]{4.818pt}{0.400pt}}
\put(198,113){\makebox(0,0)[r]{-0.1}}
\put(590.0,113.0){\rule[-0.200pt]{4.818pt}{0.400pt}}
\put(220.0,214.0){\rule[-0.200pt]{4.818pt}{0.400pt}}
\put(198,214){\makebox(0,0)[r]{-0.05}}
\put(590.0,214.0){\rule[-0.200pt]{4.818pt}{0.400pt}}
\put(220.0,315.0){\rule[-0.200pt]{4.818pt}{0.400pt}}
\put(198,315){\makebox(0,0)[r]{0}}
\put(590.0,315.0){\rule[-0.200pt]{4.818pt}{0.400pt}}
\put(220.0,416.0){\rule[-0.200pt]{4.818pt}{0.400pt}}
\put(198,416){\makebox(0,0)[r]{0.05}}
\put(590.0,416.0){\rule[-0.200pt]{4.818pt}{0.400pt}}
\put(220.0,517.0){\rule[-0.200pt]{4.818pt}{0.400pt}}
\put(198,517){\makebox(0,0)[r]{0.1}}
\put(590.0,517.0){\rule[-0.200pt]{4.818pt}{0.400pt}}
\put(269.0,113.0){\rule[-0.200pt]{0.400pt}{4.818pt}}
\put(269,68){\makebox(0,0){1}}
\put(269.0,497.0){\rule[-0.200pt]{0.400pt}{4.818pt}}
\put(366.0,113.0){\rule[-0.200pt]{0.400pt}{4.818pt}}
\put(366,68){\makebox(0,0){2}}
\put(366.0,497.0){\rule[-0.200pt]{0.400pt}{4.818pt}}
\put(464.0,113.0){\rule[-0.200pt]{0.400pt}{4.818pt}}
\put(464,68){\makebox(0,0){3}}
\put(464.0,497.0){\rule[-0.200pt]{0.400pt}{4.818pt}}
\put(561.0,113.0){\rule[-0.200pt]{0.400pt}{4.818pt}}
\put(561,68){\makebox(0,0){4}}
\put(561.0,497.0){\rule[-0.200pt]{0.400pt}{4.818pt}}
\put(220.0,113.0){\rule[-0.200pt]{93.951pt}{0.400pt}}
\put(610.0,113.0){\rule[-0.200pt]{0.400pt}{97.324pt}}
\put(220.0,517.0){\rule[-0.200pt]{93.951pt}{0.400pt}}
\put(65,315){\makebox(0,0){$G_R$}}
\put(415,23){\makebox(0,0){$d$}}
\put(220.0,113.0){\rule[-0.200pt]{0.400pt}{97.324pt}}
\put(366,356){\raisebox{-.8pt}{\makebox(0,0){$\Diamond$}}}
\put(464,276){\raisebox{-.8pt}{\makebox(0,0){$\Diamond$}}}
\put(561,184){\raisebox{-.8pt}{\makebox(0,0){$\Diamond$}}}
\put(366.0,319.0){\rule[-0.200pt]{0.400pt}{17.586pt}}
\put(356.0,319.0){\rule[-0.200pt]{4.818pt}{0.400pt}}
\put(356.0,392.0){\rule[-0.200pt]{4.818pt}{0.400pt}}
\put(464.0,229.0){\rule[-0.200pt]{0.400pt}{22.885pt}}
\put(454.0,229.0){\rule[-0.200pt]{4.818pt}{0.400pt}}
\put(454.0,324.0){\rule[-0.200pt]{4.818pt}{0.400pt}}
\put(561.0,125.0){\rule[-0.200pt]{0.400pt}{28.185pt}}
\put(551.0,125.0){\rule[-0.200pt]{4.818pt}{0.400pt}}
\put(551.0,242.0){\rule[-0.200pt]{4.818pt}{0.400pt}}
\put(316.67,506){\rule{0.400pt}{2.650pt}}
\multiput(316.17,511.50)(1.000,-5.500){2}{\rule{0.400pt}{1.325pt}}
\multiput(318.60,495.21)(0.468,-3.552){5}{\rule{0.113pt}{2.600pt}}
\multiput(317.17,500.60)(4.000,-19.604){2}{\rule{0.400pt}{1.300pt}}
\multiput(322.60,471.45)(0.468,-3.113){5}{\rule{0.113pt}{2.300pt}}
\multiput(321.17,476.23)(4.000,-17.226){2}{\rule{0.400pt}{1.150pt}}
\multiput(326.60,451.11)(0.468,-2.528){5}{\rule{0.113pt}{1.900pt}}
\multiput(325.17,455.06)(4.000,-14.056){2}{\rule{0.400pt}{0.950pt}}
\multiput(330.60,433.94)(0.468,-2.236){5}{\rule{0.113pt}{1.700pt}}
\multiput(329.17,437.47)(4.000,-12.472){2}{\rule{0.400pt}{0.850pt}}
\multiput(334.60,418.77)(0.468,-1.943){5}{\rule{0.113pt}{1.500pt}}
\multiput(333.17,421.89)(4.000,-10.887){2}{\rule{0.400pt}{0.750pt}}
\multiput(338.60,405.60)(0.468,-1.651){5}{\rule{0.113pt}{1.300pt}}
\multiput(337.17,408.30)(4.000,-9.302){2}{\rule{0.400pt}{0.650pt}}
\multiput(342.60,394.02)(0.468,-1.505){5}{\rule{0.113pt}{1.200pt}}
\multiput(341.17,396.51)(4.000,-8.509){2}{\rule{0.400pt}{0.600pt}}
\multiput(346.60,383.85)(0.468,-1.212){5}{\rule{0.113pt}{1.000pt}}
\multiput(345.17,385.92)(4.000,-6.924){2}{\rule{0.400pt}{0.500pt}}
\multiput(350.60,375.26)(0.468,-1.066){5}{\rule{0.113pt}{0.900pt}}
\multiput(349.17,377.13)(4.000,-6.132){2}{\rule{0.400pt}{0.450pt}}
\multiput(354.60,368.09)(0.468,-0.774){5}{\rule{0.113pt}{0.700pt}}
\multiput(353.17,369.55)(4.000,-4.547){2}{\rule{0.400pt}{0.350pt}}
\multiput(358.60,362.09)(0.468,-0.774){5}{\rule{0.113pt}{0.700pt}}
\multiput(357.17,363.55)(4.000,-4.547){2}{\rule{0.400pt}{0.350pt}}
\multiput(362.60,356.09)(0.468,-0.774){5}{\rule{0.113pt}{0.700pt}}
\multiput(361.17,357.55)(4.000,-4.547){2}{\rule{0.400pt}{0.350pt}}
\multiput(366.00,351.94)(0.481,-0.468){5}{\rule{0.500pt}{0.113pt}}
\multiput(366.00,352.17)(2.962,-4.000){2}{\rule{0.250pt}{0.400pt}}
\multiput(370.00,347.94)(0.481,-0.468){5}{\rule{0.500pt}{0.113pt}}
\multiput(370.00,348.17)(2.962,-4.000){2}{\rule{0.250pt}{0.400pt}}
\multiput(374.00,343.94)(0.481,-0.468){5}{\rule{0.500pt}{0.113pt}}
\multiput(374.00,344.17)(2.962,-4.000){2}{\rule{0.250pt}{0.400pt}}
\multiput(378.00,339.95)(0.685,-0.447){3}{\rule{0.633pt}{0.108pt}}
\multiput(378.00,340.17)(2.685,-3.000){2}{\rule{0.317pt}{0.400pt}}
\multiput(382.00,336.95)(0.462,-0.447){3}{\rule{0.500pt}{0.108pt}}
\multiput(382.00,337.17)(1.962,-3.000){2}{\rule{0.250pt}{0.400pt}}
\put(385,333.17){\rule{0.900pt}{0.400pt}}
\multiput(385.00,334.17)(2.132,-2.000){2}{\rule{0.450pt}{0.400pt}}
\put(389,331.17){\rule{0.900pt}{0.400pt}}
\multiput(389.00,332.17)(2.132,-2.000){2}{\rule{0.450pt}{0.400pt}}
\put(393,329.17){\rule{0.900pt}{0.400pt}}
\multiput(393.00,330.17)(2.132,-2.000){2}{\rule{0.450pt}{0.400pt}}
\put(397,327.17){\rule{0.900pt}{0.400pt}}
\multiput(397.00,328.17)(2.132,-2.000){2}{\rule{0.450pt}{0.400pt}}
\put(401,325.67){\rule{0.964pt}{0.400pt}}
\multiput(401.00,326.17)(2.000,-1.000){2}{\rule{0.482pt}{0.400pt}}
\put(405,324.67){\rule{0.964pt}{0.400pt}}
\multiput(405.00,325.17)(2.000,-1.000){2}{\rule{0.482pt}{0.400pt}}
\put(409,323.67){\rule{0.964pt}{0.400pt}}
\multiput(409.00,324.17)(2.000,-1.000){2}{\rule{0.482pt}{0.400pt}}
\put(413,322.67){\rule{0.964pt}{0.400pt}}
\multiput(413.00,323.17)(2.000,-1.000){2}{\rule{0.482pt}{0.400pt}}
\put(417,321.67){\rule{0.964pt}{0.400pt}}
\multiput(417.00,322.17)(2.000,-1.000){2}{\rule{0.482pt}{0.400pt}}
\put(421,320.67){\rule{0.964pt}{0.400pt}}
\multiput(421.00,321.17)(2.000,-1.000){2}{\rule{0.482pt}{0.400pt}}
\put(425,319.67){\rule{0.964pt}{0.400pt}}
\multiput(425.00,320.17)(2.000,-1.000){2}{\rule{0.482pt}{0.400pt}}
\put(433,318.67){\rule{0.964pt}{0.400pt}}
\multiput(433.00,319.17)(2.000,-1.000){2}{\rule{0.482pt}{0.400pt}}
\put(429.0,320.0){\rule[-0.200pt]{0.964pt}{0.400pt}}
\put(441,317.67){\rule{0.964pt}{0.400pt}}
\multiput(441.00,318.17)(2.000,-1.000){2}{\rule{0.482pt}{0.400pt}}
\put(437.0,319.0){\rule[-0.200pt]{0.964pt}{0.400pt}}
\put(452,316.67){\rule{0.964pt}{0.400pt}}
\multiput(452.00,317.17)(2.000,-1.000){2}{\rule{0.482pt}{0.400pt}}
\put(445.0,318.0){\rule[-0.200pt]{1.686pt}{0.400pt}}
\put(472,315.67){\rule{0.964pt}{0.400pt}}
\multiput(472.00,316.17)(2.000,-1.000){2}{\rule{0.482pt}{0.400pt}}
\put(456.0,317.0){\rule[-0.200pt]{3.854pt}{0.400pt}}
\put(508,314.67){\rule{0.964pt}{0.400pt}}
\multiput(508.00,315.17)(2.000,-1.000){2}{\rule{0.482pt}{0.400pt}}
\put(476.0,316.0){\rule[-0.200pt]{7.709pt}{0.400pt}}
\put(512.0,315.0){\rule[-0.200pt]{23.608pt}{0.400pt}}
\put(366,305){\raisebox{-.8pt}{\makebox(0,0){$\Box$}}}
\put(464,289){\raisebox{-.8pt}{\makebox(0,0){$\Box$}}}
\put(561,369){\raisebox{-.8pt}{\makebox(0,0){$\Box$}}}
\put(366.0,283.0){\rule[-0.200pt]{0.400pt}{10.600pt}}
\put(356.0,283.0){\rule[-0.200pt]{4.818pt}{0.400pt}}
\put(356.0,327.0){\rule[-0.200pt]{4.818pt}{0.400pt}}
\put(464.0,257.0){\rule[-0.200pt]{0.400pt}{15.418pt}}
\put(454.0,257.0){\rule[-0.200pt]{4.818pt}{0.400pt}}
\put(454.0,321.0){\rule[-0.200pt]{4.818pt}{0.400pt}}
\put(561.0,321.0){\rule[-0.200pt]{0.400pt}{22.885pt}}
\put(551.0,321.0){\rule[-0.200pt]{4.818pt}{0.400pt}}
\put(551.0,416.0){\rule[-0.200pt]{4.818pt}{0.400pt}}
\end{picture}